\documentclass[letterpaper,prc,twocolumn,showpacs,floatfix,nofootinbib,preprintnumbers,superscriptaddress,amsmath,amssymb]{revtex4-1}

\usepackage[utf8]{inputenc}
\usepackage{color}
\usepackage[dvipsnames]{xcolor}
\usepackage[colorlinks=true, urlcolor=Brown, linkcolor=RoyalBlue,
            citecolor=Green]{hyperref}
\usepackage{amsmath}           
\usepackage{amsfonts}
\usepackage{graphicx}          
\usepackage{dcolumn}           
\usepackage{bm}
\usepackage{mathtools}
\usepackage{my_acronyms}

\newcolumntype{d}[1]{D{.}{.}{#1}}
\newcommand{\mc}[1]{\multicolumn{1}{c}{#1}}
\newcommand\numberthis{\addtocounter{equation}{1}\tag{\theequation}} 
\newcommand{\gras}[1]{\boldsymbol{#1}}

\newcommand{\POUNDERS}{\textsc{pounders}}
\newcommand{\HFBTHO}{\textsc{hfbtho}}
\newcommand{\UNEDF}{\textsc{unedf}}
\newcommand{\UNEDFZERO}{\textsc{unedf0}}
\newcommand{\UNEDFONE}{\textsc{unedf1}}
\newcommand{\UNEDFTWO}{\textsc{unedf2}}
\newcommand{\BPCM}{\textsc{bpcm}}
\newcommand{\SEALL}{\textsc{seall}}

\newcommand{\enm}{E^\text{NM}}
\newcommand{\pnm}{P^\text{NM}}
\newcommand{\rhonm}{\rho_{\rm c}}
\newcommand{\knm}{K^\text{NM}}
\newcommand{\ms}{M_{s}^{*-1}}
\newcommand{\mv}{M_{v}^{*-1}}
\newcommand{\asym}{a_\text{sym}^\text{NM}}
\newcommand{\lsym}{L_\text{sym}^\text{NM}}

\makeglossaries

\begin{document}

\preprint{APS/123-QED}

\title{Microscopically-based energy density functionals for nuclei using the density matrix expansion: Full optimization and validation}


\author{R. Navarro Pérez}
\email[]{navarrop@ohio.edu}
\affiliation{Institute of Nuclear and Particle Physics and Department
  of Physics and Astronomy, Ohio University, Athens, OH 45701, USA}

\author{N. Schunck}
\email[]{schunk1@llnl.gov}
\affiliation{Nuclear and Chemical Sciences Division, Lawrence Livermore
  National Laboratory, Livermore, CA 94551, USA}

\author{A. Dyhdalo}
\email[]{dyhdalo.2@osu.edu}
\affiliation{Department of Physics, The Ohio State University,
  Columbus, OH 43210, USA}

\author{R.J Furnstahl}
\email[]{furnstahl.1@osu.edu}
\affiliation{Department of Physics, The Ohio State University,
  Columbus, OH 43210, USA}

\author{S.K. Bogner}
\email[]{bogner@nscl.msu.edu}
\affiliation{National Superconducting Cyclotron Laboratory and
  Department of Physics and Astronomy, Michigan State University, East
  Lansing, MI 48824, USA}

\date{\today}

\begin{abstract}
\begin{description}
\item[Background]
Energy density functional methods provide a generic framework to compute 
properties of atomic nuclei starting from models of nuclear potentials and the 
rules of quantum mechanics. Until now, the overwhelming majority of functionals 
have been constructed either from empirical nuclear potentials such as the 
Skyrme or Gogny forces, or from systematic gradient-like expansions in the 
spirit of the density functional theory for atoms.
\item[Purpose]
We seek to obtain a usable form of the nuclear energy density 
functional that is rooted in the modern theory of nuclear forces. We thus 
consider a functional obtained from the density matrix expansion of local nuclear 
potentials from chiral effective field theory. We propose a parametrization of 
this functional carefully calibrated and validated on selected ground-state properties that is suitable 
for large-scale calculations of nuclear properties.
\item[Methods]
Our energy functional comprises two main components. The first component is a non-local
functional of the density and corresponds to the direct part (Hartree 
term) of the expectation value of local chiral potentials on a Slater 
determinant. Contributions to the mean field and the energy of this term are 
computed by expanding the spatial, finite-range components of the chiral 
potential onto Gaussian functions. The second component is a local functional of the 
density and is obtained by applying the density matrix expansion to the 
exchange part (Fock term) of the expectation value of the local chiral 
potential. We apply the {\UNEDFTWO} optimization protocol to determine the 
coupling constants of this energy functional.
\item[Results]
We obtain a set of microscopically-constrained functionals for local chiral 
potentials from leading-order up to next-to-next-to-leading order with and 
without three-body forces and contributions from $\Delta$ excitations. These 
functionals are validated on the calculation of nuclear and neutron matter, 
nuclear mass tables, single-particle shell structure in closed-shell nuclei and 
the fission barrier of $^{240}$Pu. Quantitatively, they perform noticeable 
better than the more phenomenological Skyrme functionals.
\item[Conclusions]
The inclusion of higher-order terms in the chiral perturbation expansion seems 
to produce a systematic improvement in predicting nuclear binding energies 
while the impact on other observables is not really significant. This result is 
especially promising since all the fits have been performed at the 
single-reference level of the energy density functional approach, where 
important collective correlations such as center-of-mass correction, rotational 
correction or zero-point vibrational energies have not been taken into account 
yet.
\end{description}
\end{abstract}


\maketitle


\section{Introduction}

\gls{chiEFT} provides the framework for the modern theory of nuclear forces 
\cite{epelbaum2012,machleidt2016}. It allows the systematic construction of 
nuclear interaction potentials from first principles by introducing an expansion 
of the momentum and pion mass over the chiral symmetry breaking scale (of the order of 1 
GeV). Using chiral interactions to compute properties of heavy nuclei relevant 
to applications such as fission, nucleosynthesis or superheavy science poses a 
number of challenges. These interactions are presumed to 
represent realistic in-medium nuclear forces. Therefore, they should only be 
used in the framework of many-body methods that fully incorporate all many-body 
correlations induced by these potentials. In light nuclei, the no-core shell 
model \cite{barrett2013} or Quantum Monte-Carlo methods \cite{carlson2015} are 
popular examples of such direct approaches; in heavier nuclei, alternative 
methods such as the coupled-cluster \cite{hagen2014} or in-medium similarity 
renormalization group \cite{hergert2016} can provide good approximations of the 
exact many-body solution for nuclei near closed-shell. In spite of 
very impressive recent success, the majority of nuclei remain out of reach to 
{\it ab initio} methods, and the most microscopic approach available relies on 
the nuclear \gls{EDF} formalism \cite{bender2003}.

The \gls{EDF} approach stands in contrast to {\it ab initio} approaches in that 
it is based on enforcing that the wave-function of the nucleus take a simple 
form such as a Slater determinant in the \gls{HF} theory or a quasiparticle 
vacuum in the \gls{HFB} theory\footnote{This statement applies in the 
\gls{SR-EDF} version of the \gls{EDF} approach. In its \gls{MR-EDF} version, 
one needs to consider two different reference states.}. By definition, such an ansatz for the many-body 
wavefunction cannot be compatible with the use of realistic potentials, and 
most energy functionals are instead derived from phenomenological \gls{NN} potentials 
such as the zero-range, Skyrme potential or the  
finite-range Gogny force \cite{bender2003,stone2007}. The parameters of these 
effective \gls{NN} forces are typically adjusted on properties of nuclear 
matter or finite nuclei. Because of their phenomenological nature, these 
\gls{EDF}s do not offer a way to systematically improve their predictive power.

For these reasons, one would like to combine the predictive power, systematic 
power-counting scheme, and connection to \gls{QCD} of chiral potentials with the 
computational scaling, versatility and physics intuition of phenomenological 
potentials. One route to achieving this is based on the \gls{DME} of 
expectation values \cite{negele1972,Negele:1975zz,Gebremariam:2010ni,
Gebremariam:2009ff}. Thanks to the Wick theorem, the expectation value of an 
arbitrary potential on a product state can be expressed as a functional of the 
one-body density matrix (or generalized density if pairing correlations are 
present). In the general case, the density matrix is fully non-local, that is, 
of the form $\rho(\gras{r}\sigma\tau,\gras{r}'\sigma'\tau')$ with $\sigma$, 
$\tau$ the spin and isospin projections, respectively. The basic idea of the 
\gls{DME}  is to expand $\rho$ around the local density $\rho(\gras{r})$ in 
order to turn the expectation value of the potential into a functional of the 
local density and gradient-like corrections. 

The method was first outlined by Negele and Vautherin in \cite{negele1972,
Negele:1975zz}. Several refinements to the original method to increase its 
accuracy were proposed in \cite{Gebremariam:2010ni,Gebremariam:2009ff,
dobaczewski2010,carlsson2010}. In \cite{stoitsov2010}, the \gls{DME} was 
applied to both the direct and exchange terms of the \gls{HF} expectation value 
for unregulated momentum-space chiral potentials. The parameters of the resulting \gls{EDF} were 
adjusted approximately using the singular value decomposition algorithm, and 
were tested in calculations of radii, single-particle spectra in doubly-closed 
shell nuclei and deformation energy. While promising, the authors reported 
numerical instabilities in the practical implementation of the \gls{DME}, and 
emphasized that direct terms were treated in the local density approximation 
and that tensor contributions to the \gls{EDF} had been neglected. 

The goal of this paper is to remedy some of these limitations, 
in particular by taking into account recent developments in 
\gls{chiEFT} and employing high performance computing tools. Precisely, we want
to fully calibrate and validate an energy functional constrained by {\it local} 
chiral nuclear potentials. To this end, we build the \gls{EDF} (in the \gls{ph} 
channel) by computing the expectation value of chiral potentials on a Slater 
determinant. We use the \gls{DME} of local chiral potentials presented in 
\cite{Dyhdalo:2016tgx} to recast the exchange contribution in the form of a 
local functional of the density. We adjust the coupling constants of this 
\gls{EDF} to ground-state properties of finite nuclei by solving the \gls{HFB} 
equation. We provide results for \gls{EDF}s corresponding to different orders 
in the chiral expansion up to \gls{N2LO} order. $\Delta$ excitations and 
\gls{3N} forces are included.

The paper is organized as follows. In section II, we recall the expressions for 
\gls{NN} and \gls{3N} chiral potentials in coordinate space. Section III 
describes how these potentials can be transformed into an \gls{EDF} with the 
\gls{DME} (more details are given in \cite{Dyhdalo:2016tgx}), 
and how these \gls{EDF}s are implemented in current \gls{DFT} solvers. We also 
provide in that section the result of our calibration process. In section IV, 
we test the predictive power of these \gls{EDF}s on the equation of state of
nuclear matter, mass tables, \gls{sp} energies of doubly-closed shell nuclei 
and the fission barrier of $^{240}$Pu. Finally, we present some conclusions and 
perspectives in section V.


\section{Local Chiral Potential in Coordinate Space}

We consider local chiral potentials up to \gls{N2LO} with and without $\Delta$ 
excitations including also \gls{3N} forces~\cite{ Piarulli:2014bda,Piarulli:2016vel}. 
Chiral interactions contain finite- and zero-range contributions, with the 
zero-range couplings usually fine-tuned to reproduce low energy $\pi N$ and 
$NN$ scattering data along with selected few-body properties and sometimes
properties of
nuclei up to Oxygen \cite{Epelbaum:2014sza,Ekstrom:2015rta,Carlsson:2015vda,Piarulli:2014bda,
Piarulli:2016vel,Reinert:2017usi}. In the present work, we implement the finite-range 
contributions `as is', since they correspond to the long-range pion physics, 
which is well described by \gls{chiEFT}. In contrast, the zero-range 
contribution will be replaced by a Skyrme-like potential, and we will take the 
contact coupling constants as adjustable parameters to be determined on 
selected properties of finite nuclei.

The finite-range contributions depend on a few set of parameters
including the pion mass $m_\pi$, the $\Delta-N$ mass splitting
$M_{\Delta-N}$, the pion decay constant $f_\pi$, the nucleon axial
vector coupling $g_A$, the $N$-to-$\Delta$ axial vector coupling
$h_A$, as well as the \gls{LECs} $c_1$, $c_2$, $c_3$,
$c_4$ and the \gls{LECs} linear combination $b_3 + b_8$\footnote{For simplicity we retain 
the $b_3 + b_8$ combination even though it has been shown to be redundant \cite{Long:2011}.}. 
Although the values for most of these parameters are well determined, the \gls{LECs} have been
determined through different analyses of low energy $\pi N$ and $NN$
scattering observables yielding different results
\cite{Rentmeester:1999vw, Buettiker:1999ap, Rentmeester:2003mf,
  Entem:2003ft,PavonValderrama:2005wv, Krebs:2007rh,
  Ekstrom:2013kea,Perez:2013oba}. For this particular work, we use the
determination of Ref.~\cite{Krebs:2007rh} and leave the study of the
impact of the value of the \gls{LECs} for a future work.

Since the finite-range potentials obtained from their corresponding diagrams 
diverge as $r$ goes to zero, in practice a short range regulator is used in 
order to make scattering and structure calculations feasible. While several 
arguments have been made about the effects of the regulator on the statistical 
and systematic uncertainties~\cite{Valderrama:2008kj,PavonValderrama:2010fb,
Piarulli:2014bda}, for this work we restrict ourselves to the particular 
regulator
\begin{equation}
f(r) = \left[1 - \exp\left(-\frac{r^2}{R_c^2}\right) \right]^n,
\label{eq:regulator}
\end{equation}
with $R_c = 1.0$ fm and $n=6$. We also leave the study of the dependence on the 
strength of the regulator controlled by the $R_c$ and $n$ parameters for a 
future work. The inclusion of the regulator is one of the improvements over the 
early work of \cite{stoitsov2010}. In the two following subsections we recall the expressions for the 
long-range part only of the \gls{NN} and \gls{3N} chiral potentials. 


\subsection{Two-Body Potential}

The finite-range contribution to the local chiral potential in coordinate space 
is given by
\begin{align}
V_{\chi }(\bm{r}) 
& = 
\left( V_C(r) + W_C(r) \bm{\tau}_1 \cdot \bm{\tau}_2 \right) \nonumber \\
& + 
\left( V_S(r) + W_S(r) \bm{\tau}_1 \cdot \bm{\tau}_2 \right) \bm{\sigma}_1 \cdot \bm{\sigma}_2 \nonumber \\
& + 
\left( V_T(r) + W_T(r) \bm{\tau}_1 \cdot \bm{\tau}_2 \right) \hat{S}_{12}(\bm{\hat{r}}),
\label{eq:ChiralPotential}
\end{align}
where $r \equiv |\bm{r}|$, $\hat{S}_{12}(\bm{\hat{r}})$ is the usual tensor 
operator
\begin{equation}
\hat{S}_{12}(\bm{\hat{r}}) = 
3 (\bm{\sigma}_1\cdot\bm{\hat{r}})
  (\bm{\sigma}_2\cdot\bm{\hat{r}}) -\bm{\sigma}_1 \cdot \bm{\sigma}_2,
\end{equation}
and $\bm{\sigma}_i$ ($\bm{\tau}_i$) is the spin (isospin) operator for
the $i$th particle. The potential components at \gls{LO}
correspond to the well-known one-pion exchange and are given by
\begin{align}
W_S^{\rm LO}(r) &= \frac{m_\pi^3}{12 \pi} \left(\frac{g_A}{2 f_\pi} \right)^2 Y(r) \\
W_T^{\rm LO}(r) &= \frac{m_\pi^3}{12 \pi} \left(\frac{g_A}{2 f_\pi} \right)^2 Y(r) T(r)
\end{align}
where $Y(r)$, $U(r)$ and $T(r)$ are the usual Yukawa, scalar and tensor 
functions, respectively,
\begin{equation}
Y(r) = \frac{e^{-x}}{x}, \ 
U(r) = 1 + \frac{1}{x}, \ 
T(r) = 1 + \frac{3}{x}U(r), 
\label{eq:Yukawa}
\end{equation}
with $x = m_\pi r$. The potential components at \gls{NLO} including only nucleons and 
pions are given by
\begin{widetext}
\begin{align}
W_C^{\rm NLO}(r) 
& = + \frac{m_\pi^5}{8\pi^3(2f_\pi)^4} \frac{1}{x^4} 
\left\{ 
x \left[1+10g_A^2 -g_A^4\left(23+4x^2\right)\right]K_0(2x) 
+ \left[1+2g_A^2\left(5+2x^2\right) -g_A^4\left(23+12x^2\right) \right]K_1(2x) 
\right\}, 
\\
V_S^{\rm NLO}(r) 
& = +\frac{m_\pi^5}{2\pi^3} \left( \frac{g_A}{2f\pi} \right)^4 
\frac{1}{x^4} \Big[ 3 x K_0(2x) 
 + \big(3 +2x^2\big) K_1(2x) \Big], 
\\
V_T^{\rm NLO}(r) 
& = -\frac{m_\pi^5}{8\pi^3} \left( \frac{g_A}{2f\pi} \right)^4 
\frac{1}{x^4} \Big[12 x K_0(2x)
+ \left(15 +4x^2\right) K_1(2x) \Big],
\end{align}
\end{widetext}
where $K_0(x)$ and $K_1(x)$ are the modified Bessel functions of the second 
kind. The potential components at \gls{N2LO} including only nucleons and pions 
are given by
\begin{widetext}
\begin{align}
V_C^{\rm N2LO}(r) & = 
+\frac{3}{2}\frac{g_A^2 m_\pi^6}{(2f_\pi)^4\pi^2} \frac{e^{-2x}}{x^6}
\left[2 c_1 x^2\left(1+x\right)^2 + c_3 \left(6+12x+10x^2+4x^3+x^4 \right) \right], \\
W_S^{\rm N2LO}(r) & = +\frac{1}{3}\frac{g_A^2 m_\pi^6}{(2f_\pi)^4\pi^2} \frac{e^{-2x}}{x^6} c_4 (1+x)(3+3x+2x^2), \\
W_T^{\rm N2LO}(r) & = -\frac{1}{3}\frac{g_A^2 m_\pi^6}{(2f_\pi)^4\pi^2} \frac{e^{-2x}}{x^6} c_4 (1+x)(3+3x+x^2).
\end{align}
\end{widetext}
The expressions for the potential components coming from one- and two-$\Delta$ 
excitations can be found in the supplemental material of 
Ref.~\cite{Dyhdalo:2016tgx}.


\subsection{Three-Body Potential}

A general, local, three-body potential consists of all permutations with 
respect to the three two-body subsystems,
\begin{equation}
V^{\rm 3N} = V_{12} + V_{23} + V_{13},
\label{eq:V3N_general}
\end{equation}
where the $V_{ij}$ potential depends on two of the relative coordinates and the 
spin and isospin of the three particles,
\begin{equation}
V_{ij} = V(\bm{r}_{ik},\bm{r}_{jk},\sigma_1,\tau_1,\sigma_2,\tau_2,\sigma_3,\tau_3),
\quad \bm{r}_{ij} = \bm{r}_i - \bm{r}_j.
\end{equation}
Because of the symmetry of the potential under subscript interchange
in the \gls{DME} implementation (see Ref.~\cite{Dyhdalo:2016tgx}) only one
of the three terms in \eqref{eq:V3N_general} is necessary to express
the full three-body potential. For definiteness, we choose
$V_{23}(\bm{r}_{21},\bm{r}_{31},\{\sigma\tau\})$.

The different terms appearing in the \gls{3N} chiral potential can be 
classified as (i) long-range, which are vertices of $c_i$ or $h_A$ and have no 
Dirac delta functions, (ii) intermediate-range, which are vertices of $c_i$, 
$h_A$ or $c_D$ and have one Dirac delta function, and (iii) short-range, which 
are vertices of $c_i$, $h_A$, $c_D$ or $c_E$ with two Dirac delta functions. In 
our implementation of the \gls{3N} chiral interaction in an \gls{EDF}, we 
include only the long-range terms along with the $c_i$ and $h_A$ 
intermediate-range terms. We assume that all short-range terms and $c_D$ 
vertices can be effectively absorbed by the optimization of the contact 
couplings on nuclear properties.

There is no contribution from \gls{3N} potentials up to \gls{N2LO}, unless 
$\Delta$ excitations are included. The \gls{3N} potentials at \gls{NLO} with 
$\Delta$ and \gls{N2LO} have a very similar structure that can be summarized as
\begin{align}
V_{3N}^{\rm NLO\Delta} & = \sum_{i=1}^3 \alpha_i^{\rm NLO\Delta} V_{C,i} \\
V_{3N}^{\rm N2LO} & = \sum_{i=1}^3 \alpha_i^{\rm N2LO} V_{C,i} + V_D + V_E.
\end{align}
We employ the \gls{NLO}$\Delta$ label to emphasize that this contribution is 
only present when the $\Delta$ contributions are included. As mentioned before, 
the $V_D$ (short-range term controlled by $c_D$) and $V_E$ (short-range term 
controlled by $c_E$) terms are in fact not included in the present 
implementation. The $\alpha_i$ prefactors are given by
\begin{align}
\alpha_1^{\rm NLO\Delta} & = 0, \quad
\alpha_2^{\rm NLO\Delta}   = -\frac{h_A^2m_\pi^6g_A^2}{2592f_\pi^4\pi^2M_{\Delta-N}}, 
\nonumber\\
\alpha_3^{\rm NLO\Delta} & =  \frac{h_A^2m_\pi^6g_A^2}{10368f_\pi^4\pi^2M_{\Delta-N}},
\end{align}
and
\begin{align}
\alpha_1^{\rm N2LO} & = \frac{c_1m_\pi^6g_A^2}{16f_\pi^4\pi^2}, \quad
\alpha_2^{\rm N2LO}   = \frac{c_3m_\pi^6g_A^2}{288f_\pi^4\pi^2}, \nonumber\\
\alpha_3^{\rm N2LO} & = \frac{c_4m_\pi^6g_A^2}{576f_\pi^4\pi^2},
\end{align}
and the $V_{C,i}$ potentials are given by
\begin{widetext}
\begin{align}
V_{C,1} &= 
(\bm{\tau}_2 \cdot \bm{\tau}_3) 
(\bm{\sigma}_2 \cdot \bm{\hat{r}}_{21})
(\bm{\sigma}_3 \cdot \bm{\hat{r}}_{31}) U(r_{21}) Y(r_{21}) U(r_{31}) Y(r_{31}) 
\\
V_{C,2} &= 
(\bm{\tau}_2 \cdot \bm{\tau}_3)
\left\{ 
\frac{16\pi^2}{m_\pi^6} (\bm{\sigma}_2 \cdot \bm{\sigma}_3) \delta^3(\bm{r}_{21}) \delta^3(\bm{r}_{31})
- 
\frac{4\pi}{m_\pi^3} \left[S_{23} (\bm{\hat{r}}_{21}) T(r_{21})
+ 
(\bm{\sigma}_2 \cdot \bm{\sigma}_3) \right] Y (r_{21}) \delta^3(\bm{r}_{31}) 
\right.\nonumber\\
& 
\left.
- 
\frac{4\pi}{m_\pi^3} \left[S_{23} (\bm{\hat{r}}_{31}) T(r_{31}) 
+ 
(\bm{\sigma}_2 \cdot \bm{\sigma}_3) \right] Y (r_{31}) \delta^3(\bm{r}_{21}) 
\right. \nonumber \\
& \left. 
+ 
\left[ 
9 (\bm{\sigma}_2 \cdot \bm{\hat{r}}_{21})(\bm{\sigma}_3 \cdot \bm{\hat{r}}_{31})(\bm{\hat{r}}_{21} \cdot \bm{\hat{r}}_{31}) 
- 
3 (\bm{\sigma}_2 \cdot \bm{\hat{r}}_{21})(\bm{\sigma}_3 \cdot \bm{\hat{r}}_{21}) 
- 
3 (\bm{\sigma}_2 \cdot \bm{\hat{r}}_{31})(\bm{\sigma}_3 \cdot \bm{\hat{r}}_{31})
+ 
(\bm{\sigma}_2 \cdot \bm{\sigma}_3)  
\right]
T(r_{21}) Y(r_{21})  T(r_{31}) Y(r_{31}) 
\right.\nonumber\\
& \left.
+ (\bm{\sigma}_2 \cdot \bm{\sigma}_3) Y(r_{21}) Y(r_{31}) 
+ S_{23}(\bm{\hat{r}}_{21}) T(r_{21})Y(r_{21}) Y(r_{31}),
+ S_{23}(\bm{\hat{r}}_{31}) T(r_{31})Y(r_{31}) Y(r_{21}) \vphantom{\frac{16\pi^2}{m_\pi^6} } 
\right\} \\
V_{C,3} & = 
\bm{\tau}_2 \cdot (\bm{\tau}_3 \times \bm{\tau}_1) 
\left\{ \frac{16\pi^2}{m_\pi^6}
\delta^3(\bm{r}_{21}) \delta^3(\bm{r}_{31}) \bm{\sigma}_2 \cdot (\bm{\sigma}_3 \times \bm{\sigma}_1) 
 \right.\nonumber \\
& \left. 
-\frac{12\pi}{m_\pi^3} (\bm{\sigma}_2 \cdot \bm{\hat{r}}_{21})
\bm{\hat{r}}_{21} \cdot (\bm{\sigma}_3 \times \bm{\sigma}_1) T(r_{21}) Y(r_{21}) \delta^3(\bm{r}_{31})
+\frac{4\pi}{m_\pi^3} \bm{\sigma}_2 \cdot (\bm{\sigma}_3 \times \bm{\sigma}_1)
\frac{3}{m_\pi r_{21}} U(r_{21})) Y(r_{21}) \delta^3(\bm{r}_{31}) 
\right.\nonumber \\
& \left. 
-\frac{12\pi}{m_\pi^3} (\bm{\sigma}_3 \cdot \bm{\hat{r}}_{31})
\bm{\hat{r}}_{31} \cdot (\bm{\sigma}_1 \times \bm{\sigma}_2) T(r_{31}) Y(r_{31}) \delta^3(\bm{r}_{21})
+\frac{4\pi}{m_\pi^3}  \bm{\sigma}_2 \cdot (\bm{\sigma}_3 \times \bm{\sigma}_1)
\frac{3}{m_\pi r_{31}} U(r_{31})) Y(r_{31}) \delta^3(\bm{r}_{21}) 
\right.\nonumber \\
& \left. 
+9 (\bm{\sigma}_2 \cdot \bm{\hat{r}}_{21}) (\bm{\sigma}_3 \cdot \bm{\hat{r}}_{31})
\bm{\sigma}_1 \cdot (\bm{\hat{r}}_{21} \times \bm{\hat{r}}_{31}) T(r_{21})Y(r_{21}) T(r_{31}) Y(r_{31})
\right.\nonumber \\
& \left. 
-3 (\bm{\sigma}_2 \cdot \bm{\hat{r}}_{21})\bm{\hat{r}}_{21} \cdot (\bm{\sigma}_3 \times \bm{\sigma}_1)
T(r_{21}) \frac{3}{m_\pi r_{31}} U(r_{31})Y(r_{21})Y(r_{31}) 
\right.\nonumber \\
& \left. 
-3 (\bm{\sigma}_3 \cdot \bm{\hat{r}}_{31}) \bm{\hat{r}}_{31} \cdot (\bm{\sigma}_1 \times \bm{\sigma}_2)
T(r_{31}) \frac{3}{m_\pi r_{21}}U(r_{21})) Y(r_{21}) Y(r_{31}) 
\right.\nonumber \\
& \left. 
-\bm{\sigma}_1 \cdot (\bm{\sigma}_2 \times \bm{\sigma}_3)
\frac{3}{m_\pi r_{21}}U(r_{21})\frac{3}{m_\pi r_{31}}U(r_{31}) Y(r_{21}) Y(r_{31}) 
\vphantom{\frac{16\pi^2}{m_\pi^6}} 
\right\},
\end{align}
\end{widetext}
where the Yukawa $Y(r)$, scalar $U(r)$ and tensor $T(r)$ functions are defined 
in \eqref{eq:Yukawa}. As mentioned above, short-range terms with two Dirac 
delta functions are not included in the \gls{DME} implementation of the 
\gls{3N} interaction as we expect their effect to be absorbed by the 
calibration of the \gls{EDF} contact couplings.


\section{Implementation of $\chi$EFT in DFT}
\label{sr-edf}

In our implementation of chiral interactions in the \gls{DFT} framework, we 
will seek to write the total energy of a nucleus in the following form,
\begin{equation}
E = 
E_{\chi}^{\rm Har.}
+
E_{\chi}^{\rm Skyrme}
+
E^{\rm Coul.} + E^{\rm pair}.
\label{eq:decomposition}
\end{equation}
The first term is the Hartree (direct) contribution to the expectation value of 
the long-range part of the local chiral potentials on Slater determinants 
reference states. In practice, we will see below that only the two-body chiral 
potential contributes to it because we only work with time-even systems. The 
second term is formally identical to an extended Skyrme-like functional and 
contains {\it both} the effects of the short-range part of the chiral 
potentials (in an effective way) and of the exchange contribution from the 
long-range part (through the \gls{DME}). Finally, the terms 
$E^{\rm Cou}$ and $E^{\rm pair}$ are the usual Coulomb and pairing energy, 
obtained here by following exactly the same recipes as in \cite{kortelainen2010,
kortelainen2012,kortelainen2014}.

Two remarks are in order at this point:
\begin{itemize}
\item It should be clear from Eq.\eqref{eq:decomposition} that we restrict 
ourselves to the \gls{SR-EDF} level. In other words, we seek to calibrate a 
functional built out of a single, \gls{HFB} reference state in complete analogy 
with, e.g., the {\UNEDF} family of functionals \cite{kortelainen2010,
kortelainen2012,kortelainen2014}, the {\BPCM} functional \cite{baldo2008,
baldo2013,baldo2017} or the {\SEALL} functional \cite{refId0,Bulgac:2017bho}. As a result, 
we expect our functional to be limited in its description of the fine structure 
of $N=Z$ nuclei, for instance mirror displacement energies where isospin mixing 
and restoration are essential \cite{satula2016}) or the arc-like structure of 
binding energies near closed shell nuclei caused by quadrupole correlation 
energies \cite{bender2006,kluepfel2009}. We should also expect limitations in 
describing the shell structure of closed shell nuclei, where effects such as 
particle-vibration couplings should be taken into account \cite{colo2010}.
\item In this work, we take pairing functionals derived from a zero-range, 
surface-volume, two-body force as in the {\UNEDF} functionals. The primary 
motivation for this choice is to focus on the effect of the \gls{DME} on the 
\gls{ph} channel only before considering its application to the \gls{pp} 
channel. Early studies of pairing observables in finite nuclei with chiral 
potentials also suggest that it is mostly the short-range part of the latter 
that affect the \gls{pp} channel \cite{lesinski2008,duguet2009,hebeler2009}.
Finally, since we work at the \gls{SR-EDF} level, the consistency of the 
generating kernels between the two channels is not really an issue.
\end{itemize}

To achieve the decomposition \eqref{eq:decomposition}, we express the 
expectation value of chiral potentials on a Slater determinant reference state. 
In Section \ref{subsec:NN}, we briefly recall how this works for the two-body 
channel \eqref{eq:ChiralPotential}: the Hartree term is expanded as a sum of 
Gaussians, while the Fock term is transformed into a generalized Skyrme 
functional with the \gls{DME}. In Section \ref{subsec:NNN}, we 
give the expressions for the three-body channel, where only the Fock term
contributes; detailed derivations can be found in \cite{Dyhdalo:2016tgx}.


\subsection{Two-Body Potentials}
\label{subsec:NN}

In configuration space, the contribution to the energy from a two-body 
potential reads
\begin{equation}
E^{\rm NN} = 
\frac{1}{2} \sum_{ij} \langle ij| {\cal V}_\chi^{\rm NN} |kl \rangle
\rho_{ki}\rho_{lj},
\label{eq:ENN_conf}
\end{equation}
with $\rho_{ij}$ the matrix elements of the one-body density matrix on an 
arbitrary basis of the single-particle Hilbert space. The two-body potential is 
antisymmetrized,
\begin{equation}
{\cal V}_\chi^{\rm NN} = V_\chi^{\rm NN}(1-P_\sigma P_\tau P_r),
\end{equation}
with the usual spin-, isospin- and space-exchange operators $P_\sigma$, 
$P_\tau$ and $P_r$
\begin{equation}
P_\sigma \equiv \frac{1}{2}(1+\gras{\sigma}_1\cdot\gras{\sigma}_2), \quad \quad
 P_\tau   \equiv \frac{1}{2}(1+\gras{\tau}_1  \cdot\gras{\tau}_2  ).
\label{eq:ExchangeOperators}
\end{equation}
The antisymmetrization operator results in direct and exchange contributions, 
also referred to as the Hartree and Fock energies respectively. 

After transforming \eqref{eq:ENN_conf} to coordinate space by inserting 
resolutions of the identity, changing to relative $({\bm r})$ and center of 
mass $({\bm R})$ coordinates and assuming translational invariance along with 
a local potential, the two-body interaction energy term becomes
\begin{align}
E^{\rm NN} &= \frac{1}{2} {\rm Tr}_1 {\rm Tr}_2 \int d{\bm  R}\int d{\bm r}\;
\langle {\bm r} \sigma_1 \tau_1 \sigma_2 \tau_2 | V_\chi({\bm r}) | {\bm r} \sigma_3 \tau_3 \sigma_4 \tau_4 \rangle 
\nonumber\\
& \times 
\left[ \rho_1 \left( {\bm R} + \frac{\bm r}{2} \right) \rho_2 \left({\bm R} - \frac{\bm r}{2} \right) \right.
\nonumber\\
& \left.
-\rho_1 \left({\bm R} - \frac{\bm r}{2}, {\bm R} + \frac{\bm r}{2} \right)
 \rho_2 \left({\bm R} + \frac{\bm r}{2}, {\bm R} - \frac{\bm r}{2} \right) P_{12}^{\sigma \tau} \right],
\label{eq:FiniteRangeEnergy}
\end{align}
where the traces refer to summation over spin and isospin quantum numbers 
and the local density matrix is
\begin{equation}
  \rho(\bm{x}) \equiv \rho(\bm{x},\bm{x}).
\end{equation}
The first term in Eq.~\eqref{eq:FiniteRangeEnergy} corresponds to the Hartree 
energy, while the second one to the Fock energy. The following subsections 
describe our implementation of each of these two terms.  


\subsubsection{Hartree Term}

The one-body density matrix in Eq.~\eqref{eq:FiniteRangeEnergy} can be
decomposed into scalar-isoscalar, scalar-isovector, vector-isoscalar
and vector-isovector components \cite{bender2003}
\begin{align}
\rho(\bm{x}\sigma_1\tau_1,\bm{y}\sigma_2\tau_2) 
& = 
\frac{1}{4}
\Big[ 
\rho_0(\bm{x},\bm{y}) + \rho_1(\bm{x},\bm{y})\tau_z 
\nonumber\\
& +
\bm{S}_0(\bm{x},\bm{y}) \cdot \bm{\sigma} +
\bm{S}_1(\bm{x},\bm{y}) \cdot \bm{\sigma} \tau_z 
\Big].
\label{eq:DensityMatrixDecomposed}
\end{align}
Inserting this decomposition into the first term of 
Eq.~\eqref{eq:FiniteRangeEnergy} and performing the traces one obtains
\begin{multline}
E_{\rm H} = 
\frac{1}{2} \sum_{t=0,1} \int d \bm{r}\int d \bm{R}\; \Big[
\rho_t(\bm{R}^+) \rho_t(\bm{R}^-) \Gamma_{\rho \rho}^t \nonumber\\
  + 
\bm{S}_t(\bm{R}^+) \cdot \bm{S}_t(\bm{R}^-) \Gamma_{SS}^t \nonumber\\
 + 
(\bm{S}_t(\bm{R}^+) \cdot \hat{\bm r} ) (\bm{S}_t(\bm{R}^-) \cdot \hat{\bm r} ) \Gamma_{Sr}^t \Big],
\label{eq:ENN_Hartree}
\end{multline}
where $\bm{R}^{\pm} = \bm{R} \pm \frac{\bm r}{2}$ and
\begin{subequations}
  \begin{align*}
    \Gamma_{\rho \rho}^t &=
    \begin{dcases}
      V_C &\quad \text{for} \quad t = 0 \\
      W_C &\quad \text{for} \quad t = 1
    \end{dcases}
    \numberthis
    \\
    \Gamma_{SS}^t &=
    \begin{dcases}
      V_S - V_T &\quad \text{for} \quad t = 0 \\
      W_S - W_T &\quad \text{for} \quad t = 1
    \end{dcases}
    \numberthis
    \\
    \Gamma_{Sr}^t &=
    \begin{dcases}
      3 \; V_T &\quad \text{for} \quad t = 0 \\
      3 \; W_T &\quad \text{for} \quad t = 1
    \end{dcases}
    \numberthis
  \end{align*}
  \label{eq:hartree_functions}
\end{subequations}
with $t=0$ ($t=1$) indicating the isoscalar (isovector) case. Note that for 
systems with time-reversal symmetry, all terms diagonal in the spin density 
vanish, i.e., $\bm{S}(\bm{x}) = 0$. Hence for even-even nuclei, only the terms 
proportional to the central part of the potential in $\Gamma_{\rho \rho}^t$ 
contribute to the Hartree energy.

\begin{figure*}[!htb]
\includegraphics[width=0.9\linewidth]{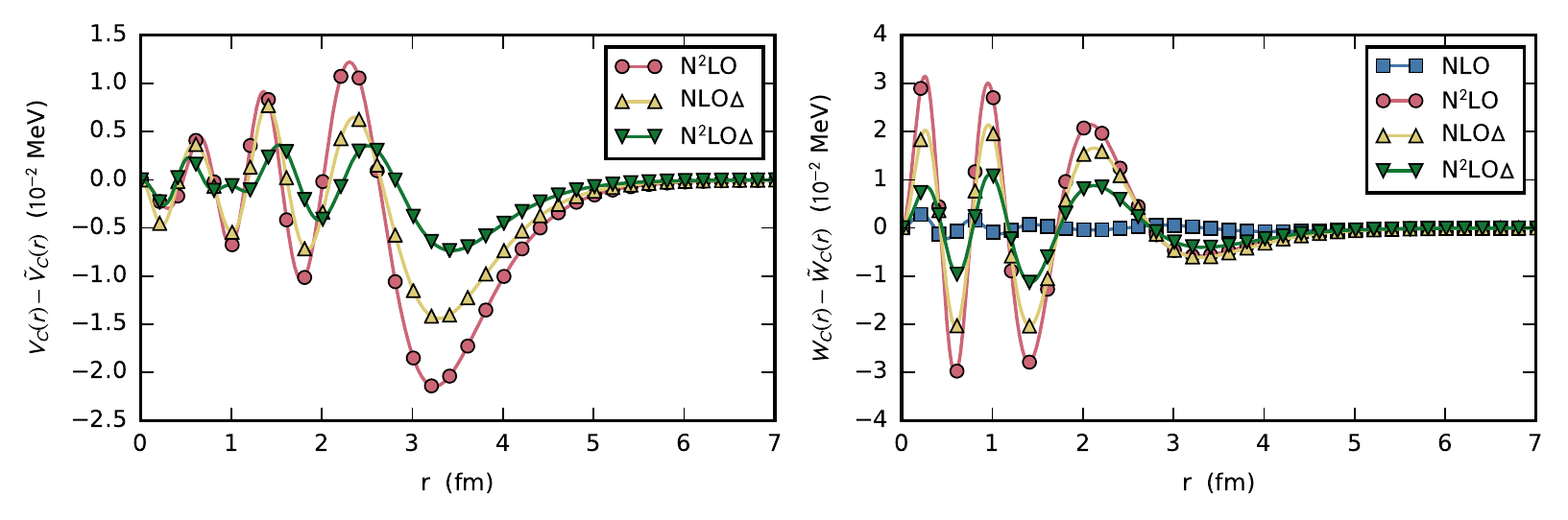}
\caption{Difference between the potentials $V_C(\gras{r})$ and $W_C(\gras{r})$ 
         up to a certain chiral order and their corresponding approximations by 
         a sum of five Gaussian functions shown in Eq.~\eqref{eq:VcGauss} and~\eqref{eq:WcGauss}. }
\label{fig:gaussians_diff}
\end{figure*}

At this stage the \gls{DME} could be applied to the calculation of the Hartree 
energy. However, it has been established that computing this term exactly 
provides a more precise description of the density fluctuations and energy 
contribution~\cite{Negele:1975zz,Sprung:1975vby}. Furthermore, the inclusion of 
the \gls{DME} approximation in the calculation of the Hartree field introduces 
numerical instabilities~\cite{Stoitsov:2010ha}. For these reasons, we choose to 
compute the direct term exactly. To compute the matrix element of the finite 
range of the chiral potential, we expand the Yukawa form factors on a series of 
Gaussian functions \cite{dobaczewski2009}. This allows us to take full advantage of the 
many analytic properties of Gaussian matrix elements in the Harmonic Oscillator basis 
\cite{Younes:2009fh} and of the existing implementation of the Gogny force in 
the latest version of {\HFBTHO} \cite{perez2017}. As we show in 
figure \ref{fig:gaussians_diff}, five Gaussian functions already give an excellent 
approximation to the spatial part of the potential.  

The chiral potential in Eq.~\eqref{eq:ChiralPotential} is expressed in a 
spin-isospin operator basis. In the code {\HFBTHO} \cite{perez2017}, 
the finite range part of the Gogny functional is implemented in a exchange 
operator basis, that is,
\begin{equation}
V_{\rm G} = \sum_{i=1}^N \big(W_i + B_i P_\sigma + H_i P_\sigma P_\tau + M_i P_\sigma P_\tau \big)
e^{-\gras{r}^2/\mu_i^2}.
\end{equation}
After inserting the definitions of Eq.~\eqref{eq:ExchangeOperators} and 
rewriting the terms in the Gogny functional, we find
\begin{align}
V_{\rm G} = \sum_{i=1}^N 
& 
\left[ W_i + \frac{B_i}{2} + \frac{H_i}{2} + \frac{M_i}{4} \right. \nonumber\\
& 
\left. + \left( \frac{B_i}{2} + \frac{M_i}{4} \right) \gras{\sigma}_1 \cdot \gras{\sigma}_2  
       + \left( \frac{H_i}{2} + \frac{M_i}{4} \right) \gras{\tau}_1 \cdot \gras{\tau}_2 
\right.
\nonumber\\
& \left.
+ \frac{M_i}{4} (\gras{\sigma}_1 \cdot \gras{\sigma}_2)(\gras{\tau}_1 \cdot \gras{\tau}_2) \right]e^{-r^2/\mu_i^2}.
\end{align}
Since only systems with time-reversal symmetry are being considered for 
this work, only the central components of the Chiral potential are considered 
in Eq.\eqref{eq:ChiralPotential}. Therefore, we can set $B_i = M_i = 0$ and use 
the approximations
\begin{align}
V_C(\gras{r}) \rightarrow 
\tilde{V}_C(\gras{r}) &= \sum_{i=1}^N \left( W_i + \frac{H_i}{2} \right) e^{-\gras{r}^2/\mu_i^2}, \label{eq:VcGauss}\\
W_C(\gras{r}) \rightarrow 
\tilde{W}_C(\gras{r}) &= \sum_{i=1}^N \frac{H_i}{2} e^{-\gras{r}^2/\mu_i^2}. \label{eq:WcGauss}
\end{align}
In order to reproduce the behavior of the regulator in Eq.~\eqref{eq:regulator} 
the conditions
\begin{equation}
  H_N = - \sum_{i=1}^{N-1} H_i, \quad \quad
  W_N = - \sum_{i=1}^{N-1} W_i,
\end{equation}
are imposed. This conditions ensure that the potentials vanish as 
$r\rightarrow 0$. The remaining free parameters $W_i$, $H_i$ and $\mu_i$ are 
adjusted numerically to reproduce the central components of the chiral 
potential at different orders. In figure \ref{fig:gaussians_diff} we show the 
difference between the approximations as a sum of five Gaussian functions and 
the corresponding chiral potential up to a certain order in the chiral 
expansion. Note that while the scale in figure \ref{fig:gaussians_diff} is
$10^{-2}$MeV the potentials have an order of magnitude, at their highest values,
of $10$ MeV.


\subsubsection{Fock Term}

Inserting Eq.~\eqref{eq:DensityMatrixDecomposed} into the second term of 
Eq.~\eqref{eq:FiniteRangeEnergy}, the two-body Fock term becomes
\begin{align}
E_{\rm F} &= -\frac{1}{2} \sum_{t=0,1} \int d \bm{r} \int d \bm{R} \left[ 
\rho_t^2(\bm{R}^+,\bm{R}^-) \Xi_{\rho \rho}^t \right. \nonumber\\ 
& \left. 
- \bm{S}_t^2(\bm{R}^+,\bm{R}^-) \Xi_{SS}^t 
-(\bm{S}_t(\bm{R}^+,\bm{R}^-) \cdot \hat{\bm r} )^2  \Xi_{Sr}^t \right],
\label{eq:fock_energy}
\end{align}
where the symmetries $\rho_t(\bm{x},\bm{y})=\rho_t(\bm{y},\bm{x})$ and
$\bm{S}_t(\bm{x},\bm{y})=-\bm{S}_t(\bm{y},\bm{x})$ for time-reversal invariant 
systems have been used and the $\Xi^t$ functions are given by
\begin{subequations}
  \begin{align*}
    \Xi_{\rho \rho}^t &=
    \begin{dcases}
      \frac{1}{4} V_C + \frac{3}{4} W_C + \frac{3}{4} V_S + \frac{9}{4} W_S \qquad \text{for} \quad t = 0 \\
      \frac{1}{4} V_C - \frac{1}{4} W_C + \frac{3}{4} V_S - \frac{3}{4} W_S \qquad \text{for} \quad t = 1
    \end{dcases}
    \label{eq:xi_pp}
    \numberthis\\
    \Xi_{SS}^t &=
    \begin{dcases}
      \frac{1}{4} V_C + \frac{3}{4} W_C - \frac{1}{4} V_S - \frac{3}{4} W_S \\
	\hspace*{2.2cm}- \frac{1}{2} V_T - \frac{3}{2} W_T \qquad \text{for} \quad t = 0 \\
      \frac{1}{4} V_C - \frac{1}{4} W_C - \frac{1}{4} V_S + \frac{1}{4} W_S \\
	\hspace*{2.2cm}- \frac{1}{2} V_T + \frac{1}{2} W_T \qquad \text{for} \quad t = 1
    \end{dcases}
    \numberthis\\
    \Xi_{Sr}^t &=
    \begin{dcases}
      \frac{3}{2} V_T + \frac{9}{2} W_T \qquad \text{for} \quad t = 0 \\
      \frac{3}{2} V_T - \frac{3}{2} W_T \qquad \text{for} \quad t = 1.
    \end{dcases}
    \numberthis
  \end{align*}
\end{subequations}
The \gls{DME} consists in expanding the non-diagonal density matrices in such manner 
that the non-locality is factorized using the following formula,
\begin{align}
\rho_t   (\bm{R}^+,\bm{R}^-) & \approx 
\sum_{n=0}^{n_{\rm max}} \Pi_{n}^{\rho} (kr) \mathcal{P}_n (\bm{R}), \\
\bm{S}_t (\bm{R}^+,\bm{R}^-) & 
\approx \sum_{m=0}^{m_{\rm max}} \Pi_{m}^{s} (kr) \mathcal{Q}_m (\bm{R})
\end{align}
where the $\Pi$ functions are specified by the \gls{DME} variant and 
$\mathcal{P}_n (\bm{R})$, $\mathcal{Q}_m (\bm{R})$ denote various local 
densities. The arbitrary momentum scale $k$ in the $\Pi$ functions sets the 
scale for the fall-off in the off-diagonal direction. In this work, we follow 
common practice and truncate the expansion at $n_{\rm max} = 2$ and 
$m_{\rm max} = 1$ such that
\begin{align}
\rho_t(\bm{R}^+,\bm{R}^-) & 
\approx \Pi_0^\rho(k_{\rm F}r) \rho_t(\bm{R}) \nonumber\\
& + \frac{r^2}{6} \Pi_2^\rho
  \left[\frac{1}{4} \Delta \rho_t(\bm{R}) - \tau_t(\bm{R}) + \frac{3}{5} k_{\rm F}^2 \rho_t(\bm{R}) \right], \label{eq:dme_matter_density} \\
  S_{t,b}(\bm{R}^+,\bm{R}^-) & \approx i \Pi_1^s (k_{\rm F} r) \sum_{a=x}^z r_a J_{t,ab}(\bm{R}) \label{eq:dme_spin_density},
\end{align}
where the kinetic density $\tau_t$ and spin current density $\bm{J}_t$ are defined as
\begin{align}
\tau_t(\bm{r}) & = \nabla \cdot \nabla' \rho_t(\bm{r},\bm{r}')|_{\bm{r}=\bm{r}'}, \label{eq:tau} \\
J_{t,ab}(\bm{r}) & = -\frac{i}{2}(\nabla_a - \nabla_a')S_{t,b}(\bm{r},\bm{r}')|_{\bm{r}=\bm{r}'}  \label{eq:current}.
\end{align}
In Gebremariam's improved phase-space-averaging \gls{DME} variant
\cite{Gebremariam:2010ni,Gebremariam:2009ff}, the momentum scale $k$ is 
chosen to be the Fermi momentum $k_{\rm F}$ with the $\Pi$ functions given by
\begin{equation}
\Pi_0^{\rho} (k_{\rm F} r) 
= \Pi_2^{\rho} (k_{\rm F} r) 
= \Pi_1^s (k_{\rm F} r) 
= 3 \frac{j_1 (k_{\rm F} r)}{k_{\rm F} r }
\end{equation}
where $j_1$ is a spherical Bessel function of the first kind and $k_{\rm F}$ is 
related to the isoscalar density in the usual way,
\begin{equation}
k_{\rm F} = \left( \frac{3 \pi^2}{2} \rho_0 (\bm{R}) \right)^{1/3}.
\end{equation}
By inserting the expansions of 
Eqs.~\eqref{eq:dme_matter_density}-\eqref{eq:dme_spin_density} into the exact 
Fock energy of Eq.~\eqref{eq:fock_energy}, the Fock energy can be approximated 
by expressions involving only products of local densities. Terms beyond 
second-order in the density expansions are dropped e.g., 
$\Pi_2^\rho(k_{\rm F}r) \Pi_2^\rho(k_{\rm  F}r)$. After performing the \gls{DME} and 
organizing the different terms in Eq.~\eqref{eq:fock_energy} by the different 
densities, we find the more compact expression
\begin{align}
E_{\rm F} \approx &
\sum_{t=0,1} \int d \bm{R}\; \left[ 
g_t^{\rho\rho} \rho_t \rho_t + g_t^{\rho\tau} \rho_t \tau_t + g_t^{\rho\Delta\rho} \rho_t \Delta\rho_t \right.
\nonumber\\
& \left. + g_t^{JJ,1} J_{t,aa} J_{t,bb} + g_t^{JJ,2} J_{t,ab}
             J_{t,ab} + g_t^{JJ,3} J_{t,ab} J_{t,ba} \right]
\label{eq:ENN_Fock}
\end{align}
where the $\bm{R}$-dependence of the local densities and couplings $g$ has been 
omitted for simplicity and the coupling functions are given by
\begin{subequations}
\begin{align*}
g^{\rho \rho}_t(\rho_0) & = 
    \begin{array}[t]{ll}\displaystyle
    -\frac{4\pi}{2} \int dr r^2 \Big[ \Pi_0^\rho(k_{\rm F}r)^2 
    \nonumber\\ \displaystyle
    + \frac{r^2 k_{\rm F}^2}{5} \Pi_0^\rho(k_{\rm F}r) \Pi_2^\rho(k_{\rm F}r) \Big] \Xi_{\rho \rho}^t(r), 
   \end{array}
\label{eq:grr}
\numberthis\\
g^{\rho \tau}_t(\rho_0) & = \frac{4\pi}{2} \int dr r^2 \left[ 
  \frac{r^2}{3} \Pi_0^\rho(k_{\rm F}r) \Pi_2^\rho(k_{\rm F}) \right] \Xi_{\rho \rho}^t(r),
\label{eq:grt}
\numberthis\\
g^{\rho \Delta \rho}_t(\rho_0) &= -\frac{4\pi}{2} \int drr^2 \left[ 
  \frac{r^2}{12} \Pi_0^\rho(k_{\rm F}r) \Pi_2^\rho(k_{\rm F}) \right] \Xi_{\rho \rho}^t(r),
\label{eq:grd}
\numberthis\\
g^{JJ,1}_t(\rho_0) &= -\frac{4\pi}{2} \int dr r^2 \left[ 
  \frac{r^2}{15} \Pi_1^S(k_{\rm F}r)^2 \right] \Xi_{Sr}^t(r),
\label{eq:gJJ1}
\numberthis\\
g^{JJ,2}_t(\rho_0) &= -\frac{4\pi}{2} \int dr r^2 \left[ 
  \frac{r^2}{15} \Pi_1^S(k_{\rm F}r)^2 \right] \left[ 5 \Xi_{SS}^t(r) + \Xi_{Sr}^t(r) \right],
\label{eq:gJJ2}
\numberthis\\
g^{JJ,3}_t(\rho_0) &= -\frac{4\pi}{2} \int dr r^2 \left[ 
  \frac{r^2}{15} \Pi_1^S(k_{\rm F}r)^2 \right] \Xi_{Sr}^t(r).
\label{eq:gJJ3}
\numberthis
\end{align*}
\end{subequations}

As already highlighted in \cite{stoitsov2010}, one of the practical differences 
between Skyrme and \gls{DME}-based functionals is that each Skyrme coupling constant 
becomes a coupling function, which is dependent on the isoscalar density. The 
calculation of these density-dependent couplings requires performing several
multidimensional numerical integrals, some of them converging slowly at small 
values of $\rho_0$. However, these couplings are completely independent of the 
system being calculated or any other characteristic of the \gls{HFB} simulation like 
the basis size or oscillator length. Therefore, we can tabulate all the 
relevant coupling functions for different values of the density $\rho_0$. In 
the actual \gls{HFB} calculation the couplings are approximated via the interpolating function

\begin{figure*}[!htb]
\includegraphics[width=0.8\linewidth]{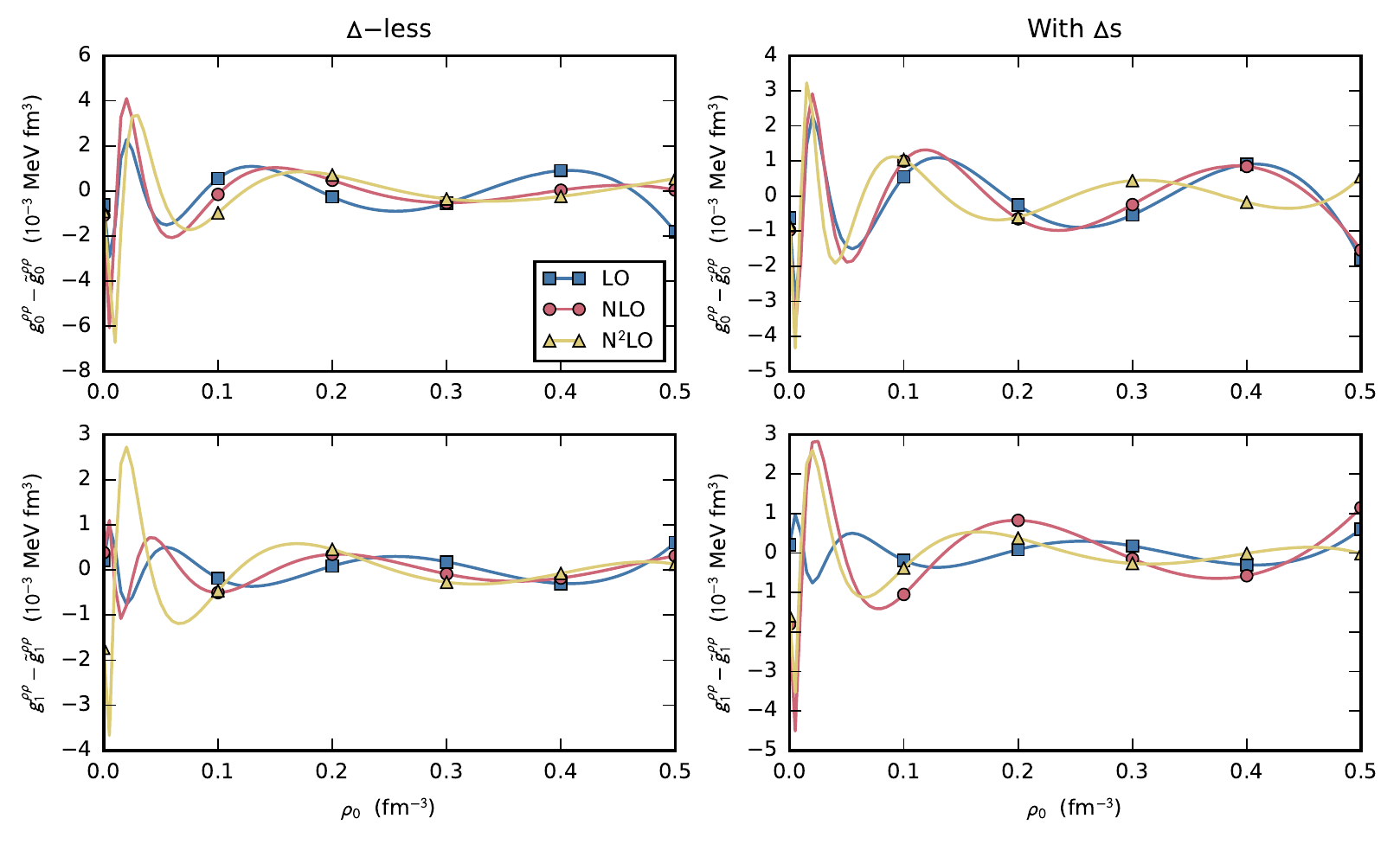}
\caption{Difference between the two-body density dependent couplings
         $g_t^{\rho\rho}$ and their corresponding interpolating function
         $\tilde{g}_t^{\rho\rho}$ given by 
         Eq.~\eqref{eq:interpolated_coupling}.}
\label{fig:couplings_diff_2n}
\end{figure*}

\begin{equation}
\tilde{g}^{uv}(\rho) = g^{uv}(0) + \sum_{i=1}^{N} a_i \tan^{-1}(b_i \rho_0^{c_i}).
\label{eq:interpolated_coupling}
\end{equation}

The parameters of the interpolating functions were adjusted to reproduce
the tabulated values using $N=3$. While other forms of interpolating
functions were considered, this one gave a better description of the
coupling functions while avoiding numerically unstable behavior at
small and large values of $\rho_0$.  An additional advantage of using
interpolating functions is that the inclusion of $\Delta$ excitations
and \gls{3N} forces does not imply any increase on computational cost
since the same type of interpolating function is used for all
cases. Figure \ref{fig:couplings_diff_2n} shows the numerical
precision of these interpolating functions for two-body
couplings. While the numerical precision of the interpolation is of the order 
of $10^{-3}$ MeV fm$^{-3}$, this is 4 orders of magnitude smaller than the 
scale of the couplings themselves, which are $\propto 10$ MeV fm$^{-3}$.


\subsection{Three-Body Term}
\label{subsec:NNN}

The contribution of the $V_{\chi}^{\rm 3N}$ three-body chiral potential to the 
total energy is given by
\begin{equation}
E^{\rm 3N} = 
\frac{1}{6} \sum_{ijk} 
\langle ijk | V_\chi^{\rm 3N} \mathcal{A}_{123} | lmn \rangle
\rho_{li}\rho_{mj}\rho_{nk},
\end{equation}
with $\mathcal{A}_{123} = (1+P_{13} P_{12} + P_{23} P_{12})(1-P_{12})$ the full 
three-body antisymmetrization operator. Since all the terms in the \gls{3N} 
Hartree energy contain at least one spin density matrix, which vanishes in 
time-reversal invariant systems, there is in fact no contribution from the 
\gls{3N} channel to the Hartree energy. The application of the \gls{DME} to the 
Fock term results in 23 trilinears of local densities, each one with its 
corresponding density-dependent coupling,
\begin{widetext}
  \begin{align}
    E_{\rm F}^{\rm 3N} & \approx \int d\bm{R} \left(
    g^{\rho_0^3}\rho_0^3 
    + g^{\rho_0^2\tau_0}\rho_0^2\tau_0 +
    g^{\rho_0^2\Delta\rho_0}\rho_0^2\Delta\rho_0 
    + g^{\rho_0(\nabla\rho_0)^2}\rho_0\nabla\rho_0\cdot\nabla\rho_0 
    + g^{\rho_0\rho_1^2}\rho_0\rho_1^2 
    + g^{\rho_1^2\tau_0}\rho_1^2\tau_0 
    + g^{\rho_1^2\Delta\rho_0}\rho_1^2\Delta\rho_0 
    \right. \nonumber \\
    & + g^{\rho_0\rho_1\tau_1}\rho_0\rho_1\tau_1
    + g^{\rho_0\rho_1\Delta\rho_1}\rho_0\rho_1\Delta\rho_1 
    + g^{\rho_0(\nabla\rho_1)^2}\rho_0\nabla\rho_1\cdot\nabla\rho_1 
    + \rho_0 \epsilon_{ijk} \left[
      g^{\rho_0\nabla\rho_0J_0}\nabla_i\rho_0J_{0,jk} + g^{\rho_0\nabla\rho_1J_1}\nabla_i\rho_1J_{1,jk} 
      \right] 
    \nonumber \\
    & + \rho_1 \epsilon_{ijk} \left[
      g^{\rho_1\nabla\rho_1J_0}\nabla_i\rho_1J_{0,jk} + g^{\rho_1\nabla\rho_0J_1}\nabla_i\rho_0J_{1,jk} 
      \right]
    + \rho_0 \left[
      g^{\rho_0J_0^2,1}J_{0,aa}J_{0,bb} + g^{\rho_0J_0^2,2}J_{0,ab}J_{0,ab} + g^{\rho_0J_0^2,3}J_{0,ab}J_{0,ba} 
      \right] 
    \nonumber \\
    & + \rho_0 \left[
      g^{\rho_0J_1^2,1}J_{1,aa}J_{1,bb} + g^{\rho_0J_1^2,2}J_{1,ab}J_{1,ab} + g^{\rho_0J_1^2,3}J_{1,ab}J_{1,ba} 
      \right] .
    \nonumber  \\
    & + \rho_1 \left. \left[
      g^{\rho_0J_0J_1,1}J_{1,aa}J_{0,bb} + g^{\rho_0J_0J_1,2}J_{1,ab}J_{0,ab} + g^{\rho_0J_0J_1,3}J_{1,ab}J_{0,ba} \right]
    \vphantom{g^{\rho_0^3}\rho_0^3} \right). 
\label{eq:ENNN_Fock}
  \end{align}
\end{widetext}

\begin{figure*}[!htb]
\includegraphics[width=0.8\linewidth]{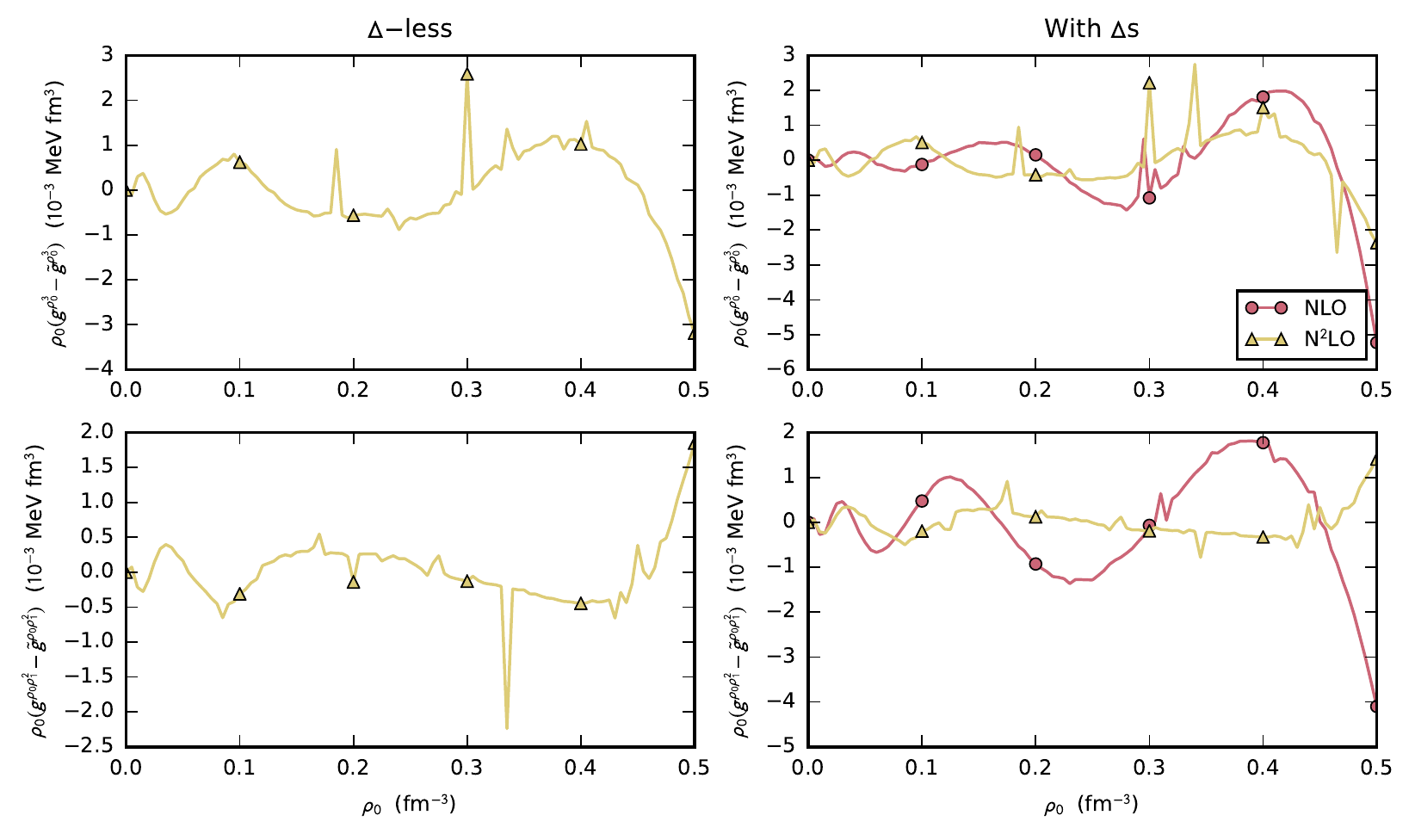}
\caption{Difference between the three-body density-dependent couplings 
         $g^{\rho_0^3}$, $g^{\rho_0\rho_1^2}$ and their corresponding 
         interpolating functions $\tilde{g}^{\rho_0^3}$, 
         $\tilde{g}^{\rho_0\rho_1^2}$ given by 
         Eq.~\eqref{eq:interpolated_coupling}. The irregularities in the
         curves show that the interpolation has an accuracy similar to the
         numerical multidimensional integral.}
\label{fig:couplings_diff_3n}
\end{figure*}

We refer the reader to~\cite{Dyhdalo:2016tgx} and its supplemental material for 
an in-depth derivation of the \gls{3N} energy Fock term and complete expressions for 
the density-dependent couplings. Similarly to the two-body case, the 
calculation of the density functions in terms of the density $\rho_0$ requires 
several multidimensional numerical integrals with slow convergence. To avoid 
calculating these coupling functions at every iteration of the \gls{HFB} 
calculation, we employ the same type of interpolating function as in 
Eq.~\eqref{eq:interpolated_coupling} and adjust its parameters to reproduce 
tabulated couplings. Figure \ref{fig:couplings_diff_3n} shows the accuracy of 
these interpolations at different chiral orders for 2 representative functions, 
$g^{\rho_0^3}$ and $g^{\rho_0\rho_1^2}$. The accuracy of the interpolation is 
comparable for all coupling functions. The irregularities in the curves show 
that the interpolation has an accuracy similar to the numerical multidimensional 
integral. As with two-body couplings, the numerical error of the interpolation 
is about 4 orders of magnitude smaller than the couplings themselves.



\subsection{Optimization of Contact Couplings}

Unlike {\it ab initio} methods in which many-body correlations are encoded into 
the nucleon wave-functions, \gls{DFT} assumes independent (quasi-)particles through 
uncorrelated, product wave-functions. Therefore, many-body correlations have to 
be included both through the form and parameters of the \gls{EDF} (single- or 
multi-reference) and through the symmetry-breaking mechanism. Recall that in 
our decomposition \eqref{eq:decomposition} of the total energy, the Skyrme-like 
part is the sum of three terms: an effective two-body Skyrme functional that 
mocks up the effects of the short range part of the chiral potentials; a 
generalized Skyrme functional of the form \eqref{eq:ENN_Fock} that contains the 
exchange contribution of the two-body chiral potential; and a generalized 
Skyrme functional of the form \eqref{eq:ENNN_Fock} for the exchange 
contribution of the \gls{3N} channel. The two-body part thus reads
\begin{multline}
E_{\rm NN}^{\rm Skyrme} = \sum_{t=0,1} \int d \bm{R} \left[ 
\left( 
U_{t0}^{\rho \rho} + 
U_{tD}^{\rho\rho} \rho_0^\gamma 
\right) \rho_t^2 + 
U_t^{\rho\tau} \rho_t \tau_t \right.\\ 
\left. + 
U_t^{\rho\Delta\rho} \rho_t \Delta \rho_t +
U_t^{\rho \nabla J} \rho_t \nabla \cdot \bm{J}_t + 
U_t^{JJ} J_{t,ab} J_{t,ab}
\right] 
\label{eq:SkyrmeEnergy}
\end{multline}
where each coupling function is given by 
\begin{equation}
U_{t}^{uu'}(\rho) = C_{t}^{uu'} + g_{t}^{uu'}\big[\rho_{0}(\gras{R})\big]
\label{eq:Ufunc}
\end{equation}
with the functions $g_{t}^{uu'}\big[\rho_{0}(\gras{R})\big]$ listed in 
Eqs.~\eqref{eq:grr}-\eqref{eq:gJJ3}. The pairing energy is given by
\begin{equation}
E^{\rm pair} = \frac{1}{4} \sum_{q=n,p} \int d \bm{R}\; 
V_0^q \left[ 1 - \frac{1}{2} \frac{\rho_0(\gras{R})}{\rhonm} \right] \tilde{\rho}^2(\gras{R}) 
\label{eq:PairingEnergy}
\end{equation}
where $\tilde{\rho}(\bm{R})$ is the pairing density and $\rhonm =0.16$ 
fm$^{-3}$. The coefficients $C_{t}^{uu'}$ and $V_0^q$ are the unknown 
parameters that we will determine in the calibration process. 

In practice, note that we are only fitting parameters in the $NN$ exchange and 
pairing channel of the functional. The two-body Hartree term, 
Eq.~\eqref{eq:ENN_Hartree}, is computed ``exactly'' (that is, without any 
adjustment of parameters) and so is the three-body Fock term, 
Eq.~\eqref{eq:ENNN_Fock}. 

For the optimization of the contact couplings, we follow the same prescription 
as for the {\UNEDFTWO} parametrization of the Skyrme functional 
\cite{kortelainen2014}. Among the fourteen parameters in 
Eqs.~\eqref{eq:SkyrmeEnergy} and \eqref{eq:PairingEnergy}, $C_{t0}^{\rho\rho}$, 
$C_{tD}^{\rho\rho}$, $C_t^{\rho\tau}$ and $\gamma$ are volume couplings and, 
therefore, can be directly related to \gls{INM} properties, which allows using 
tighter, physically-motivated bounds; see Section IV.A-C and Appendix C in 
\cite{stoitsov2010} for actual expressions relating \gls{INM} properties with 
the coupling functions \eqref{eq:Ufunc}. In practice, we thus optimize the 
following \gls{INM} parameters: $\rhonm$, $\enm$, $\knm$, $\ms$, $\asym$ and 
$\lsym$. As with all {\UNEDF} parametrizations of the Skyrme functional, we do 
not optimize the vector effective mass $\mv$ but instead keep it fixed at its 
SLy4 value, $\mv = 1.249$. The remaining eight parameters are fitted directly 
with the same bounds as for {\UNEDFTWO}.

Note that, in contrast to Skyrme \gls{EDF}s, our \gls{DME} \gls{EDF}s contain a 
finite-range term that contributes to \gls{INM} properties. The contribution of 
this finite-range term to the energy per particle for asymmetric nuclear matter 
reads 
\cite{Sellahewa:2014nia}
\begin{widetext}
\begin{equation}
  e(\rho,\beta) = \frac{1}{2} \sum_{i=1}^N \left\{ A_0^i \rho + A_1^i \rho \beta^2  + B_{nn}^i \left[\frac{1+\beta}{2} g(\mu_i k_F^n) + \frac{1-\beta}{2} g(\mu_i k_F^p)  \right] + B_{np}^i h(\mu_i k_F^n,\mu_i k_F^p) \right\},
\end{equation}
\end{widetext}
where $\rho = \rho_n + \rho_p$ is the total density, $\beta=(\rho_n-\rho_p)/\rho$ the isospin 
asymmetry parameter, $k_F^n = k_F(1+\beta)^{1/3}$ and 
$k_F^p = k_F(1-\beta)^{1/3}$ are the Fermi momenta of the corresponding
isospin symmetric system, the parameters $A$ and $B$ are 
\begin{align}
  A_0^i &= \frac{\pi^{3/2}\mu_i^3}{4} \left( 4 W_i + 2 B_i - 2 H_i - M_i\right),\\
  A_1^i &= \frac{\pi^{3/2}\mu_i^3}{4} \left( - 2 H_i - M_i\right), \\
  B_{nn}^i &= - \frac{1}{\sqrt{\pi}} \left(W_i + 2B_i- H_i -2M_i \right), \\
  B_{np}^i &=   \frac{1}{\sqrt{\pi}} \left(H_i  + 2M_i \right),
\end{align}
and the functions functions $g(q)$ and $h(q_1,q_2)$ are the result of a double 
integration of the exchange matrix elements over the same Fermi surface for 
$g$, and two different surfaces for $h$, which result in
\begin{align}
  g(q) &= \frac{2}{q^3} - \frac{3}{q} - \left(\frac{2}{q^3} - \frac{1}{q} \right) e^{-q^2} + \sqrt{\pi}{\rm erf}(q) \\
  h(q_1,q_2) &= 2 \frac{q_1^2 -q_1q_2+q_2^2-2}{q_1^3+q_2^3} e^{-\frac{(q_1+q_2)^2}{4}} \nonumber \\
  &- 2  \frac{q_1^2 +q_1q_2+q_2^2-2}{q_1^3+q_2^3} e^{-\frac{(q_1-q_2)^2}{4}}  \\
  & -\sqrt{\pi} \frac{q_1^3-q_2^3}{q_1^3+q_2^3}{\rm erf}\left(\frac{q_1-q_2}{2}\right) + \sqrt{\pi}{\rm erf}\left(\frac{q_1+q_2}{2}\right) \nonumber
\end{align}

\begin{table*}[!htb]
\caption{Parameters of the various \gls{EDF}s optimized in this work. The last 
line gives the value of the objective function at convergence of the 
optimization process. $\rhonm$ is in fm$^{-3}$; $\enm$, $\knm$, $\asym$ and 
$\lsym$ are in MeV; $\ms$ is dimensionless; $C_t^{\rho\Delta\rho}$, 
$C_t^{\rho\nabla J}$ and $C_t^J$ are in MeV fm$^5$; and $V_0^n$ and $V_0^p$ are 
in MeV fm$^3$. \label{tab:EDF}}
\begin{ruledtabular}
\begin{tabular}{l d{4.4} d{4.4} d{4.4} d{4.4} d{4.4} d{4.4} d{4.4} d{4.4}}
      & \mc{LO}  & \mc{NLO} & \mc{N2LO} & \mc{N2LO+3N} & \mc{NLO$\Delta$} & \mc{NLO$\Delta$+3N} & \mc{N2LO$\Delta$} & \mc{N2LO$\Delta$+3N} \\
\hline
      $\rhonm$               &   0.1544 &   0.1550 &   0.1584 &   0.1530 &   0.1586 &   0.1539 &   0.1556 &   0.1527 \\
      $\enm$                 & -15.8025 & -15.8000 & -15.8353 & -15.8133 & -15.8617 & -15.8135 & -15.8385 & -15.8417 \\
      $\knm$                 & 258.6536 & 254.9564 & 221.2413 & 250.3559 & 223.0304 & 250.0137 & 248.2058 & 259.2423 \\
      $\asym$                &  30.0578 &  30.5201 &  29.7554 &  29.2640 &  30.5042 &  29.6921 &  29.6983 &  30.4040 \\
      $\lsym$                &  41.9577 &  42.9947 &  40.0000 &  40.2500 &  44.2077 &  40.0000 &  41.9412 &  40.0000 \\
      $\ms$                  &   0.9763 &   0.9000 &   0.9048 &   0.9000 &   0.9000 &   0.9000 &   0.9000 &   0.9000 \\
      $\gamma$               &   0.5404 &   0.5238 &   0.3526 &   0.4931 &   0.3452 &   0.4711 &   0.4728 &   0.5301 \\
      $C_0^{\rho\Delta\rho}$ & -36.1843 & -52.6447 &   6.9892 & -29.1293 & -13.7870 & -33.2312 &   1.0440 & -35.3682 \\
      $C_1^{\rho\Delta\rho}$ & -70.2703 & -61.1454 & -65.9052 & -51.6414 & -68.9016 & -48.5797 & -69.1111 &  -4.1573 \\
      $V_0^n$                &-194.7660 &-163.9659 &-164.2242 &-164.1451 &-163.3901 &-164.7202 &-163.1916 &-163.5773 \\
      $V_0^p$                &-227.1298 &-189.0806 &-191.1035 &-190.0664 &-188.7028 &-189.6970 &-189.0264 &-190.2702 \\
      $C_0^{\rho\nabla J}$   & -62.0154 & -62.1937 & -67.1676 & -72.8370 & -94.9189 & -75.8855 & -64.0646 & -69.6724 \\
      $C_1^{\rho\nabla J}$   & -81.2615 &-104.3616 & -71.2548 & -66.7901 & -40.3346 & -46.2612 & -37.6989 & -64.9767 \\
      $C_0^J$                &-101.8945 & -84.8842 &-100.4869 &-104.9008 & -18.7145 & -86.8496 &-112.1157 &-115.9108 \\
      $C_1^J$                &  34.4538 &  31.0748 & -41.5981 & -11.0683 &  39.6320 &  14.9317 & -10.3164 & -20.7663 \\
\hline
      $f(x)$                 & 229.6620 & 164.1405 & 165.6655 & 167.6812 & 160.9050 & 165.2818 & 159.6821 & 162.6206
\end{tabular}
\end{ruledtabular}
\end{table*}

The calculation of the \gls{INM} properties simply requires performing a Taylor 
expansion around the saturation density $\rhonm$ which yields
\begin{align}
  e(\rho,\beta) &= e(\rho) + S_2(\rho) \beta^2 + S_4(\rho) \beta^4 + \ldots,\\
  e(\rho) &= \frac{\enm}{A} + \frac{\pnm}{\rhonm} \delta \rho  + \frac{\knm}{18\rhonm^2}(\delta \rho)^2 + \ldots, \\
  S_2(\rho) &= \asym + \frac{\lsym}{3\rhonm} + \delta \rho + \ldots.
\end{align}
For detailed derivations of the finite-range contributions to the infinite 
nuclear matter properties see \cite{Sellahewa:2014nia}.

The new \gls{EDF}s were optimized from \gls{LO} to \gls{N2LO}. The inclusion of 
a $\Delta$ excitation produces two additional versions of the functional at 
\gls{NLO} and \gls{N2LO}, denoted as \gls{NLO}$\Delta$ and \gls{N2LO}$\Delta$. 
Finally the incorporation of \gls{3N} forces at the appropriate orders adds three 
more versions denoted as \gls{N2LO}+\gls{3N}, \gls{NLO}$\Delta$+\gls{3N} and 
\gls{N2LO}$\Delta$+\gls{3N}. Each of these 8 \gls{EDF}s has its own set of 
Gaussian functions to represent the finite-range contribution to the Hartree 
field, density-dependent couplings for the two- and three-body Fock fields and 
calibrated two-body contact couplings to recover many-body correlations.

As already highlighted in \cite{stoitsov2010}, the parameter space of \gls{DME} 
\gls{EDF}s could be significantly different from that of traditional Skyrme 
\gls{EDF}s. To avoid possible difficulties during the optimization process 
(which was initialized with the {\UNEDFTWO} parameter set), we took advantage 
of the built-in regulator \eqref{eq:regulator}: for large values of $R_c$, 
$f(r) \rightarrow 0$ and the finite-range contributions vanish. Therefore, the 
\gls{EDF} reduces to a traditional Skyrme \gls{EDF}. Starting from the 
{\UNEDFTWO} parameter set, we thus produced intermediate parametrizations of 
all the \gls{DME} \gls{EDF}s at $R_c = 2.0$ fm. As mentioned earlier, the final 
parametrizations were obtained with $R_c = 1.0$ fm. 

It is important to note that the $\Delta$-less and $\Delta$-full version of the 
chiral potential employ different sets of low-energy constants. In this work we 
derive the \gls{EDF}s from the local potentials in coordinate space as 
presented in~\cite{Piarulli:2014bda}, for which the \gls{DME} approximation has 
been applied in~\cite{Dyhdalo:2016tgx}. The values for the \gls{LECs} and other 
physical parameters used for the chiral potentials in this work are listed in 
Table I of~\cite{Dyhdalo:2016tgx}.

The actual parameter sets of all 8 \gls{DME} \gls{EDF}s are listed in 
Table~\ref{tab:EDF}. The optimization was carried out with the {\POUNDERS} 
optimization package from Argonne National Laboratory, with all \gls{HFB} 
calculations performed with the {\HFBTHO} solver with the exact same basis 
characteristics as in \cite{kortelainen2014}. We notice that for most \gls{DME}
\gls{EDF}s, the scalar effective mass ends up at its bound. As expected, there 
are substantial variations among the different parametrizations, in particular 
when it comes to $\gamma$, $C_0^{\rho\Delta\rho}$ and $C_1^{JJ}$. Large 
fluctuations in the isovector channel are not surprising, since it is widely 
believed that the lack of constraints on these parameters comes from a lack 
experimental data in very neutron-rich nuclei. The observed fluctuations in the 
power of the density-dependence and the isoscalar surface terms are indicative 
of the strong non-linearity of the optimization process. With the exception of 
the \gls{EDF} at LO, the value of the objective function is similar for all 
\gls{EDF} at around 164$\pm$5, but we will see in the next section that there 
are significant differences in predictive powers.


\section{Validation Against Experimental Data}


\subsection{Infinite nuclear matter properties}

\begin{figure*}
\includegraphics[width=0.8\linewidth]{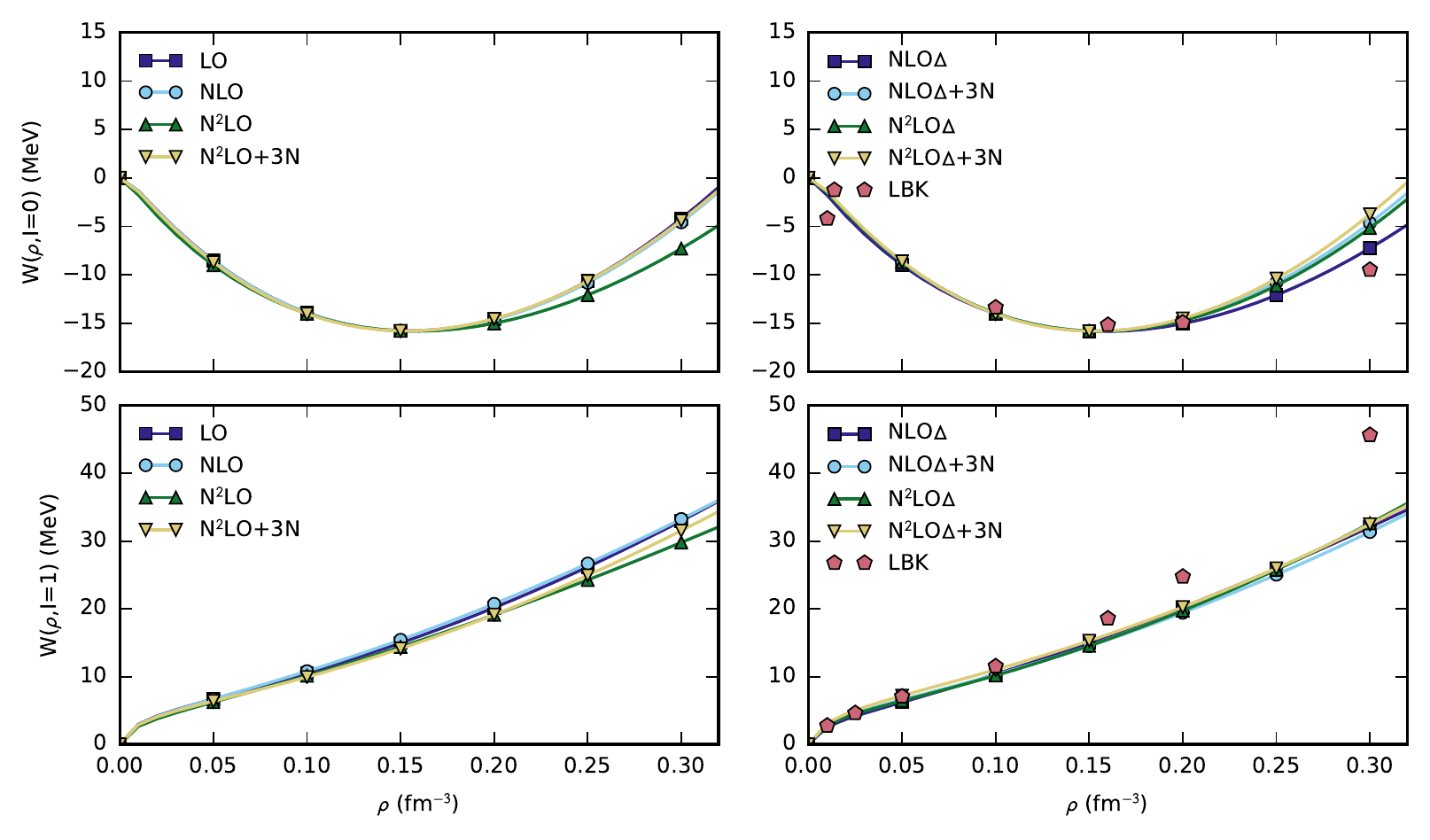}
\caption{Symmetric nuclear matter (top panels) and pure neutron matter 
         (bottom panels) for the new \gls{DME} functionals. For comparison we 
         mark with red pentagons the calculation by Logoteta, Bombaci and 
         Kievsky (LBK) for a local chiral potential with Delta isobar, labeled 
         as N3LO$\Delta$+\gls{N2LO}$\Delta$1 in~\cite{Logoteta:2016nzc}. }
\label{fig:NuclearMatter}
\end{figure*}

We calculated the \gls{EOS} with each of the new microscopically constrained 
\gls{EDF}s. As mentioned in the previous section, the inclusion of 
density-dependent couplings and finite-range contributions in the density 
functional brings additional terms to the corresponding \gls{EOS}. These terms 
were included following the derivations in~\cite{Stoitsov:2010ha, 
Sellahewa:2014nia}. In Fig.~\ref{fig:NuclearMatter} we show the energy per 
nucleon $E/A$ as a function of the density $\rho$ for \gls{SNM} and \gls{PNM}.

Since the value of the saturation density and other \gls{INM} properties at 
saturation were used to constrain the contact couplings for the \gls{EDF}s, it 
is not surprising that all curves exhibit very similar behavior around the 
saturation point. The curves for the different \gls{EDF}s start to deviate from 
one another at large values of $\rho$, specially for cases in which the 
\gls{3N} terms are not included even though the corresponding diagrams are 
present at such order. For the $\Delta$-less implementation (left panels), a 
convergence pattern can be seen when including the \gls{3N} terms, i.e., the 
difference between \gls{LO} and \gls{NLO} is larger than the difference between 
\gls{NLO} and \gls{N2LO}+\gls{3N}. Unfortunately such a convergence pattern can 
not be found in the $\Delta$-full implementation.

As a reference point we include the recent calculation by Logoteta, Bombaci and 
Kievsky (LBK) for a local chiral potential with $\Delta$ 
isobar~\cite{Logoteta:2016nzc}. While a direct comparison can not be done with 
our current results since the LBK calculations correspond to a $\Delta$-full 
implementation of the 2-body force at \gls{N3LO} and the \gls{3N} terms at 
\gls{N2LO}, it still provides a useful reference. Also for comparison, the recent
auxiliary field diffusion Monte Carlo simulations \cite{Tews:2015ufa} using a
chiral interaction at \gls{N2LO} give an energy per nucleon for \gls{SNM} between
12 and 16 MeV at saturation density depending on the value of the regulator. These
simulations are in agreement with our calculations. Overall, our parametrization 
seems to yield a stiffer \gls{EOS} than fully microscopic calculations.


\subsection{Nuclear Mass Tables}

For each of the \gls{EDF} listed in Table~\ref{tab:EDF}, we computed the 
binding energies of all even-even nuclei from $Z=8$ to $Z=120$. The driplines 
were identified by the requirement that the two-neutron separation energy 
change sign and become negative. For each even-even nucleus, we considered 11 
different configurations characterized by their axial quadrupole deformation 
$\beta_2 = -0.25, \dots, +0.25$ (by steps of 0.05). For each configuration, we 
used the small-deformation approximation of the quadrupole moment, 
$Q_2 \approx \beta_2 \sqrt{5/\pi}(Z+N)^{5/3}/100$ (in barns), to impose a 
constraint on $Q_2$ for the \gls{HFB} solution. The constraint was only active 
during the first 20 iterations of the self-consistent loop and was then 
automatically released. The binding energy retained for the even-even nucleus 
is then the lowest energy of these 11 configurations.

\begin{figure*}[!htb]
\includegraphics[width=0.45\textwidth]{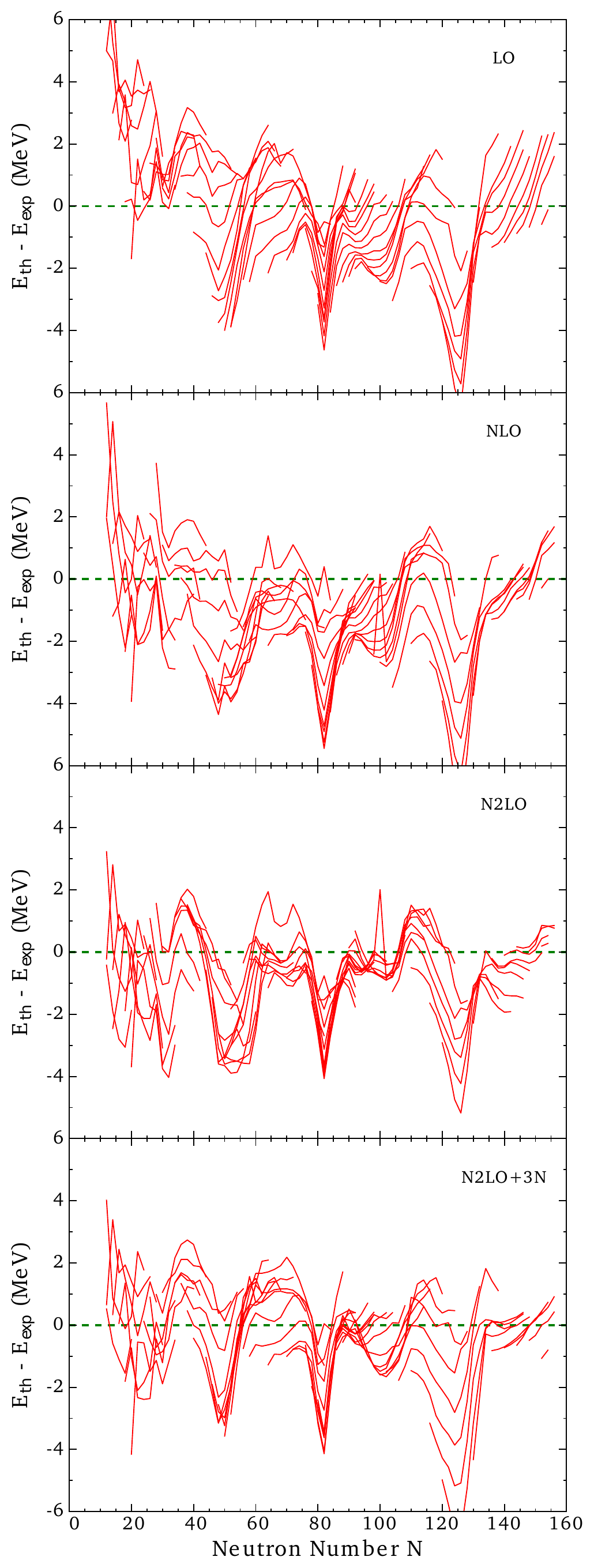}
\includegraphics[width=0.45\textwidth]{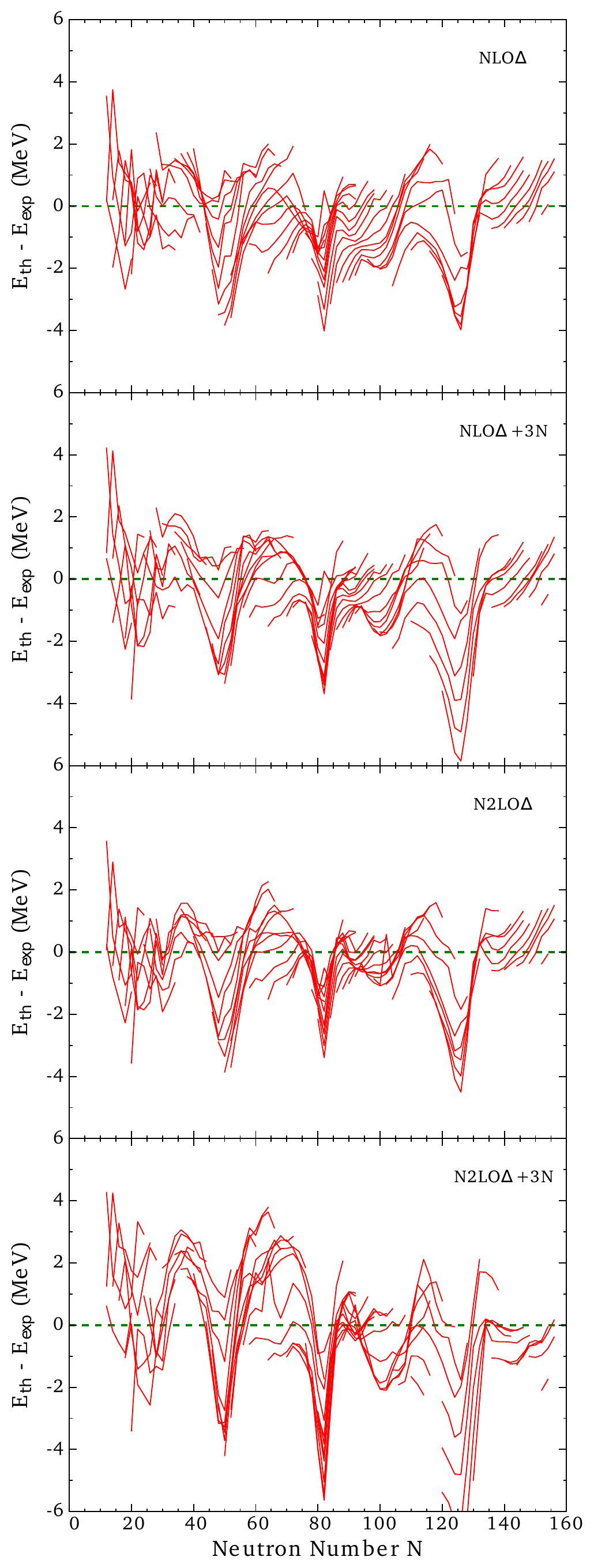}
\caption{Deviations between theoretical and experimental binding energies for 
         even-even nuclei. Experimental binding energies are extracted from the 
         2016 mass evaluation \cite{Huang2016, Wang2016} and only actual 
         measured values are used. Left panel: \gls{EDF}s without $\Delta$ 
         excitations; right panel: \gls{EDF}s with $\Delta$ excitations.}
\label{fig:masses}
\end{figure*}

Binding energies of odd-even and odd-odd nuclei were not computed explicitly, 
as it would require performing numerous blocking calculations. There are at 
least two reasons why such an explicit calculation is not mandated here: (i) 
the main focus of this work is a global assessment of microscopically-constrained 
\gls{EDF}s, not the production of a mass model, and (ii) our optimization protocol, 
also used for the Skyrme {\UNEDFTWO} functional, does not put special emphasis 
on nuclear masses, which are 1 of 5 different types of observables. For these 
reasons, we rely instead on a popular approximation, where the energy of an odd-even nucleus 
$(Z-1,N)$ (with both $Z$ and $N$ even-even) is given by
\begin{multline}
E(Z-1,N) = 
\frac{1}{2} \big[ E(Z,N) + E(Z-2,N) \big] \\
+ 
\frac{1}{2} \big[ \Delta_p(Z,N) + \Delta_p(Z-2,N) \big],
\end{multline}
with $\Delta_p(Z,N)$ the average proton pairing gap (obtained as 
$\Delta(Z,N) = \frac{1}{2}\text{Tr} \Delta\rho_p$, where $\Delta$ is the 
pairing field of the \gls{HFB} matrix). Similar formula hold for even-odd 
nuclei. For odd-odd systems, we first compute
\begin{equation}
\Delta_p(Z,N-1) = \frac{1}{2} \big[ \Delta_p(Z,N) + \Delta_p(Z,N-2) \big]
\end{equation}
and, similarly, $\Delta_p(Z-2,N-1)$, and combine them to get
\begin{multline}
E(Z-1,N-1) = 
\frac{1}{2}\big[ E(Z,N-1) + E(Z-2,N-1) \big] \\
+ 
\frac{1}{2} \big[ \Delta_p(Z,N-1) + \Delta_p(Z-2,N-1) \big]
\end{multline}
All calculations were performed with the code {\HFBTHO} in a deformed (stretched) 
basis of 20 shells with an axial deformation $\beta = \beta_2$.

\begin{table}[!htb]
\caption{\gls{rms} deviations between experimental and theoretical binding 
energies. Experimental values are taken from the 2016 Atomic Mass Evaluation 
\cite{Huang2016,Wang2016}; see text for additional details.}
\begin{ruledtabular}
\begin{tabular}{ccc}
      \gls{EDF}              & \gls{rms} & Number of nuclei \\
      \hline                
 {\UNEDFTWO}                 & 1.98 & 620 \\
   \gls{LO}                  & 1.99 & 617 \\
  \gls{NLO}                  & 2.02 & 617 \\
 \gls{N2LO}                  & 1.57 & 616 \\
 \gls{N2LO}+\gls{3N}         & 1.58 & 613 \\
  \gls{NLO}$\Delta$          & 1.41 & 618 \\
  \gls{NLO}$\Delta$+\gls{3N} & 1.46 & 617 \\
 \gls{N2LO}$\Delta$          & 1.26 & 615 \\
 \gls{N2LO}$\Delta$+\gls{3N} & 1.72 & 617 \\
\end{tabular}
\end{ruledtabular}
\label{table:masses}
\end{table}

Table~\ref{table:masses} summarizes the characteristics of the mass tables for 
each \gls{EDF}. It lists the \gls{rms} deviation between theoretical and 
experimental nuclear binding energies as well as the number of experimental 
measurements. Experimental atomic masses are taken from the 2016 Atomic Mass 
Evaluation \cite{Huang2016,Wang2016}. Nuclear binding energies are obtained 
after taking into account the binding energy of the electrons. Following 
\cite{Wang2016}, we adopt the following empirical formula
\begin{equation}
B_e(Z) = 1.44381\times 10^{-5}Z^{2.39} + 1.55468\times 10^{-12} Z^{5.35}
\end{equation}
with the energy given in MeV. We only included true experimental measurements 
and did not take into account evaluated masses. Further details on how nuclear 
binding energies are extracted from the mass evaluation can be found in 
\cite{kortelainen2010}.

Perhaps the most surprising (and promising) result is the relatively large 
variation of the results, with a \gls{rms} ranging from 1.26 MeV for 
\gls{N2LO}$\Delta$ to 2.02 MeV for \gls{NLO} -- a 60\% difference in predictive 
power. It is also very encouraging to note that the \gls{EDF}s seem to perform 
better and better overall as we go from \gls{LO} to \gls{NLO} to \gls{N2LO} and 
add $\Delta$ excitations. In fact, the quality of the \gls{N2LO}+$\Delta$ 
\gls{EDF} is rather spectacular. Without any ``beyond mean-field'' corrections 
such as the rotational or vibrational corrections, Wigner energy, etc., this 
\gls{EDF} does markedly better than {\UNEDFTWO}. Remember that (i) the 
determination of the \gls{EDF} parameters was made with the exact same protocol 
and optimizer (ii) all mass tables were computed with the exact same code, 
basis characteristics and overall algorithms to identify driplines. Therefore, 
the origin of all differences listed in Table~\ref{table:masses} can be 
attributed to the form of the \gls{EDF} only.

A visual representation of the difference between theory and experiment 
highlights a few additional interesting features of these \gls{EDF}s. For light 
nuclei, Fig.~\ref{fig:masses} shows that the \gls{LO} and \gls{NLO} \gls{EDF}s 
behave like the {\UNEDFTWO} (and older {\UNEDFONE}) \gls{EDF}: discrepancies 
with experimental masses are larger, which was explained in 
\cite{kortelainen2012,kortelainen2014} as resulting from neglecting the 
center-of-mass correction in the \gls{EDF} -- a choice that we also made for 
all the \gls{DME} \gls{EDF}s. Surprisingly, this feature is much attenuated for 
\gls{EDF}s based on higher-order chiral potentials. 

We also notice that both \gls{EDF}s including the effect of three-body force 
and, to a lesser extent, that of the $\Delta$ excitations have more pronounced 
spikes near closed-shell nuclei, as shown by comparing, e.g., the mass tables 
for \gls{N2LO} and \gls{N2LO}+\gls{3N}, or \gls{N2LO}$\Delta$ and 
\gls{N2LO}$\Delta$+\gls{3N}. Overall, we also notice that the effect of the 
three-body force seems to be the largest near closed shells, in particular near 
$^{208}$Pb.

\begin{table}[!htb]
\caption{Mean and standard deviations of the binding energy residuals for each 
         of 8 \gls{EDF}s considered in this work.}
\begin{ruledtabular}
\begin{tabular}{ccc}
      \gls{EDF}              &   mean &   $\sigma$ \\
      \hline                              
 {\UNEDFTWO}                 & -0.277 & 1.960 \\
   \gls{LO}                  & -0.288 & 1.971 \\
  \gls{NLO}                  & -1.144 & 1.659 \\
 \gls{N2LO}                  & -0.799 & 1.351 \\
 \gls{N2LO}+\gls{3N}         & -0.411 & 1.521 \\
  \gls{NLO}$\Delta$          & -0.480 & 1.327 \\
  \gls{NLO}$\Delta$+\gls{3N} & -0.461 & 1.386 \\
 \gls{N2LO}$\Delta$          & -0.343 & 1.209 \\
 \gls{N2LO}$\Delta$+\gls{3N} & -0.538 & 1.636 \\
\end{tabular}
\end{ruledtabular}
\label{table:histograms}
\end{table}

Table \ref{table:histograms} completes the picture by showing the mean value 
and standard deviations computed from the residuals of nuclear binding 
energies. Compared with {\UNEDFTWO} and \gls{LO} (which is not much different 
from {\UNEDFTWO} by construction), \gls{DME} functionals have a larger 
systematic bias -- which also tends to decrease as we go to higher order in the 
\gls{chiEFT} expansion. Conversely, the standard deviation for the \gls{DME} 
functionals is much smaller than for Skyrme, and the trend is also towards 
smaller standard deviations. Recall that for a random variable with mean $\mu$ 
and standard deviation $\sigma$, we have ${\rm rms}^2 = \sigma^2 + \mu^2$. 

\begin{figure}[!htb]
\includegraphics[width=0.45\textwidth]{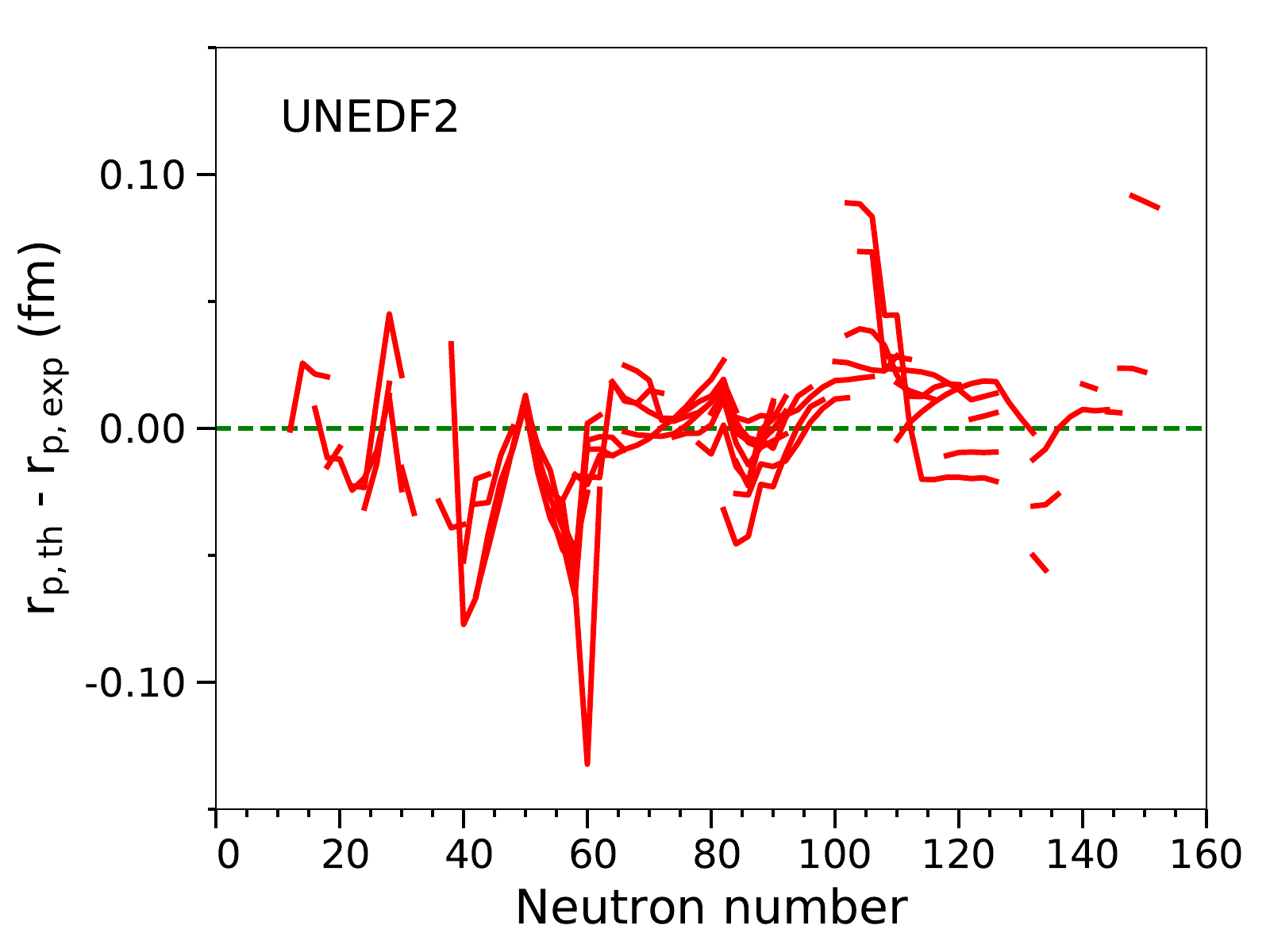}
\includegraphics[width=0.45\textwidth]{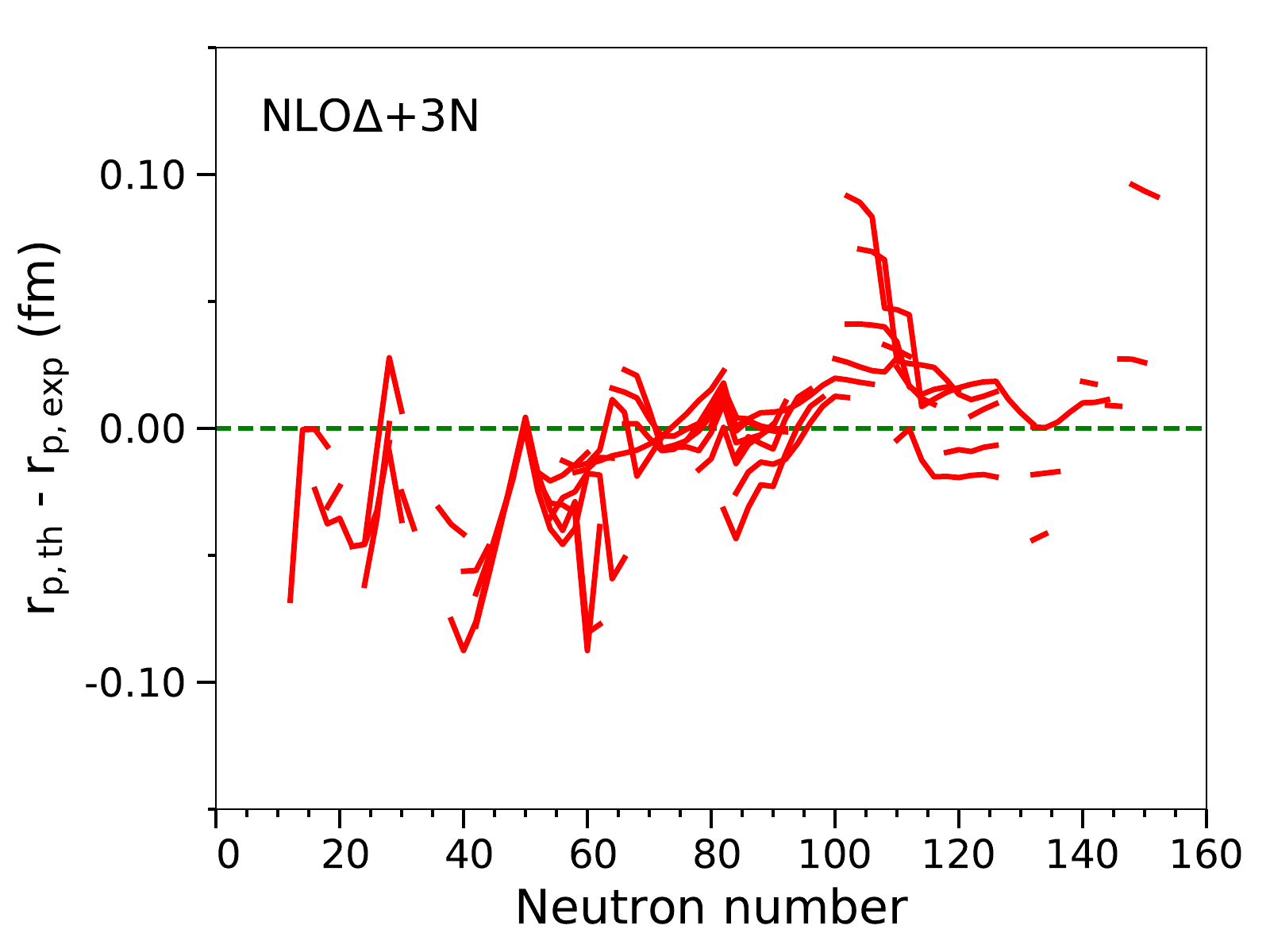}
\caption{Proton radii residuals for the {\UNEDFTWO} (top panel) and 
         \gls{NLO}$\Delta$+\gls{3N} (bottom panel) functionals. Experimental data are 
         taken from \cite{angeli2013}.}
\label{fig:radii}
\end{figure}

Another indicator of the global quality of a \gls{EDF} is the proton radius. We 
show in Fig.\ref{fig:radii} the residuals for proton radii for the 339 nuclei 
listed in \cite{angeli2013} in the two particular case of the {\UNEDFTWO} and 
the \gls{NLO}$\Delta$+\gls{3N} functionals. We extracted experimental proton 
radii from the table of \cite{angeli2013} by using the formula
\begin{equation}
r_{\rm ch}^{2} = \langle r_{p}^{2} \rangle + r_{p}^{2} + \frac{N}{Z}r_{n}^{2}
\end{equation}
with $r_{\rm ch}^{2}$ the charge radius of the nucleus, 
$\langle r_{p}^{2} \rangle$ the expectation value of the proton radius on the 
\gls{HFB} solution, and $r_{n}^{2} = -0.1161$ fm$^{2}$ and $r_{p}^{2} = 0.769$ 
fm$^{2}$ the charge mean square radii of the free neutron and proton, 
respectively. Table \ref{table:radii} lists the \gls{rms}, mean value and 
standard deviation of the residuals for all 8 functionals listed in Table 
\ref{tab:EDF}. Overall, the prediction of proton radii is on par with competing 
functionals, see, e.g., \cite{baldo2013,goriely2013}, although it is slightly 
worse than for the Skyrme {\UNEDFTWO}. We also observe a similar effect as for 
masses: \gls{DME} functionals have a larger systematic bias than the pure 
Skyrme {\UNEDFTWO}. However, this bias is very small ($<0.01$ fm) and may not 
be very significant.

\begin{table}[!htb]
\caption{\gls{rms}, mean and standard deviations, in fm, for the proton radius 
         residuals for each of 8 \gls{EDF}s considered in this work. }
\begin{ruledtabular}
\begin{tabular}{cccc}
      \gls{EDF}              &   rms    &   mean   & $\sigma$ \\
      \hline                                       
 {\UNEDFTWO}                 & 0.027449 & +0.000373 & 0.027487 \\
   \gls{LO}                  & 0.028710 & -0.000465 & 0.028749 \\
  \gls{NLO}                  & 0.033489 & -0.003573 & 0.033347 \\
 \gls{N2LO}                  & 0.033558 & -0.002725 & 0.033496 \\
 \gls{N2LO}+\gls{3N}         & 0.035225 & +0.001547 & 0.035243 \\
  \gls{NLO}$\Delta$          & 0.028432 & -0.006602 & 0.027695 \\
  \gls{NLO}$\Delta$+\gls{3N} & 0.031102 & -0.003854 & 0.030908 \\
 \gls{N2LO}$\Delta$          & 0.029878 & -0.004295 & 0.029611 \\
 \gls{N2LO}$\Delta$+\gls{3N} & 0.032645 & -0.000004 & 0.032694 \\
\end{tabular}
\end{ruledtabular}
\label{table:radii}
\end{table}


\subsection{Shell Structure}

We turn to the \gls{sp} shell structure of closed shell nuclei. As a reminder, 
we extract \gls{sp} energies of the nucleus $(Z,N)$ by performing blocking 
calculations \cite{ring2000} at the equal filling approximation \cite{perez-martin2008} 
in the neighboring odd nuclei, e.g., $(Z,N\pm 1)$ for neutrons \gls{sp} states; 
see, e.g., \cite{rutz1998,rutz1999,duguet2001,duguet2001-a,schunck2010,
tarpanov2014} for studies of the blocking prescription on the ground-state 
properties of odd nuclei. Specifically, we define 
\begin{subequations}
\begin{eqnarray}
E_{\rm s.p.}^{(\text{part.})} & = & E_{\rm bl}(A+1) - E(A), \label{eq:spEpart} \\
E_{\rm s.p.}^{(\text{hole})}  & = & E(A) - E_{\rm bl}(A-1), \label{eq:spEhole}
\end{eqnarray}
\end{subequations}
where $A$ is the particle number of the reference, doubly-magic, nucleus of 
interest and $E_{\rm bl}$ is the energy of the blocked configuration in the
neighboring odd nucleus. The labels ``hole'' and ``particle'' refer to whether
the corresponding \gls{sp} levels would be, respectively, fully occupied or 
empty in the corresponding \gls{HF} calculation of the doubly-magic nucleus. 

This method presents two advantages. First, it ensures the consistency of the 
calculations for all observables. Whether we consider masses, \gls{sp} energies 
or fission barriers, we always perform computations in the same \gls{HFB} framework 
with the Lipkin-Nogami correction. Second, we automatically include the small 
shape polarization induced by the blocking calculation -- even though this 
polarization is restricted here to axial shapes owing to the built-in 
symmetries of {\HFBTHO} \cite{schunck2010,tarpanov2014}.

We recall that in {\HFBTHO}, blocking configurations can only be specified by the 
Nilsson quantum numbers $[Nn_z\Lambda]\Omega$ of the requested \gls{sp} state; 
see \cite{stoitsov2013} for details. Since these quantum numbers are only valid 
approximately ($\Omega$ corresponds to a conserved symmetry of the mean field, 
but not the others; see \cite{bohr1975} for a discussion), the convergence of 
the blocking calculations can sometimes fail. In particular, we found that for 
low-j orbitals, we had to introduce tiny constraints either on the expectation 
value of $\hat{Q}_2$ or $\hat{Q}_4$ in order to converge the blocking 
calculations. Since the effective \gls{sp} energy is defined as an energy 
difference, the numerical error introduced is very small -- less than 50 keV 
overall. Note that similar difficulties were experienced with the {\UNEDF} 
family of \gls{EDF}s presented in \cite{kortelainen2010,kortelainen2012,
kortelainen2014}.

\begin{figure}[!htb]
\includegraphics[width=0.45\textwidth]{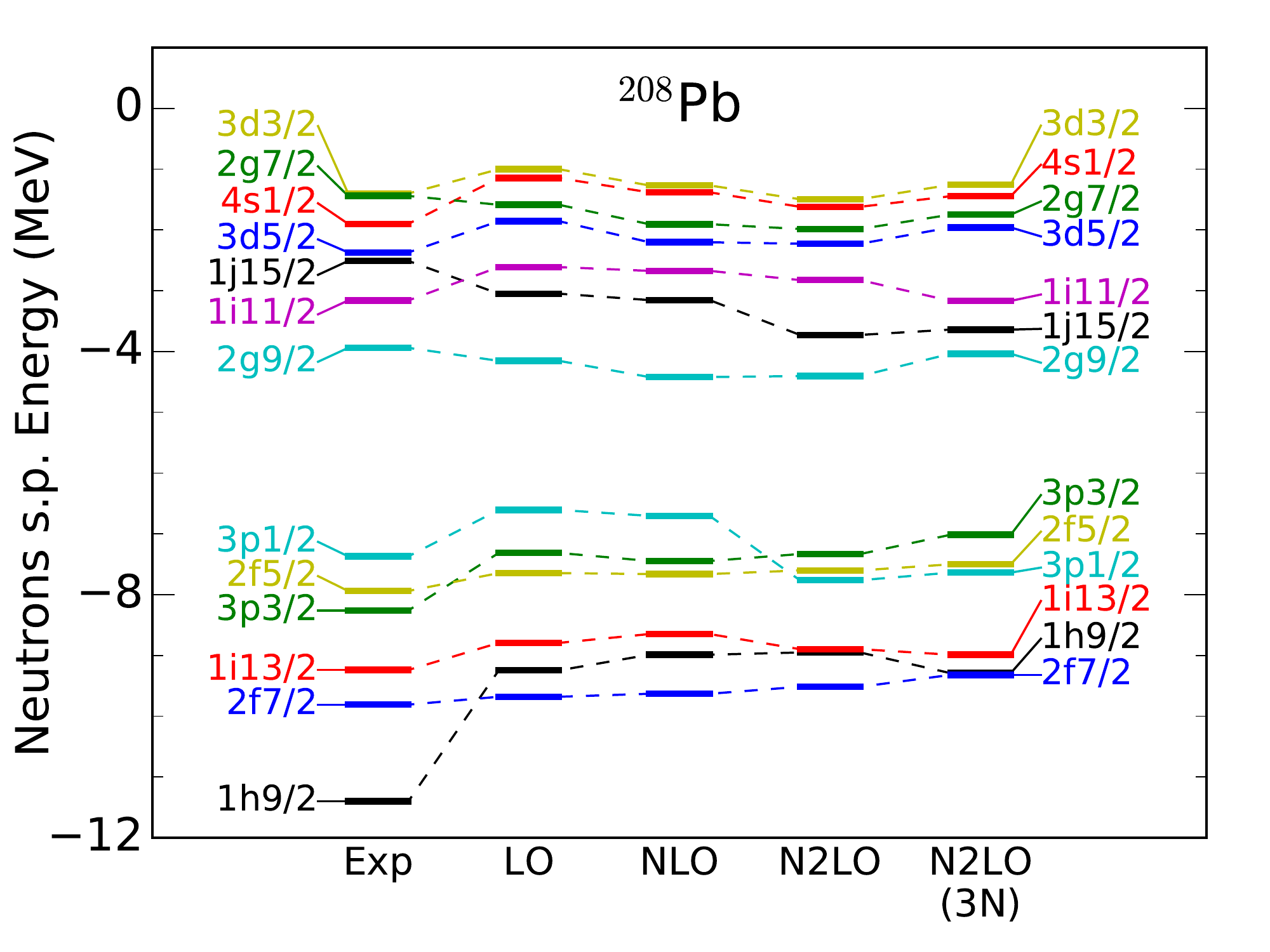}
\includegraphics[width=0.45\textwidth]{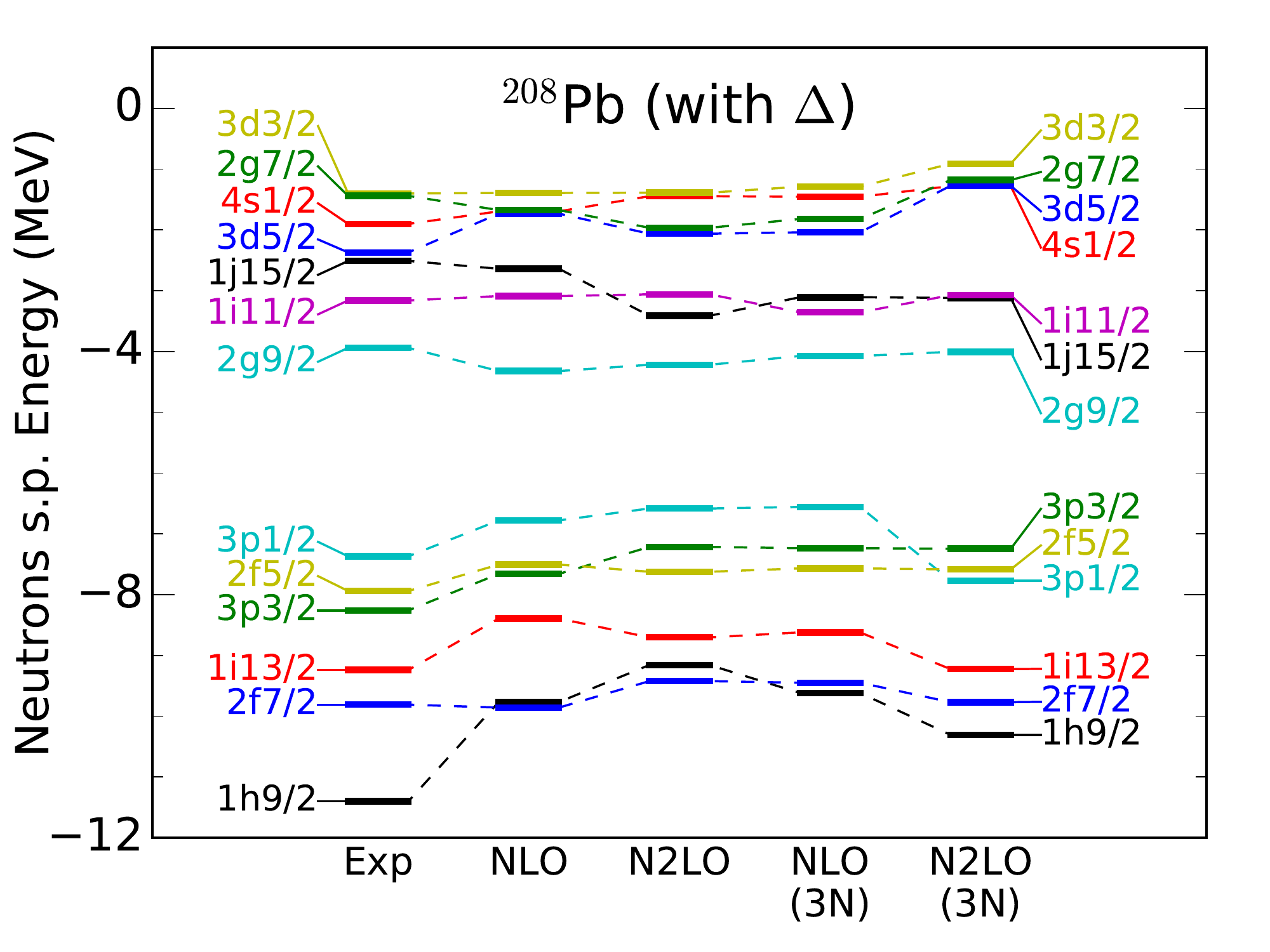}
\caption{Neutrons \gls{sp} levels in $^{208}$Pb extracted from blocking 
         calculations. In the top panel, \gls{EDF}s derived from \gls{NN} and 
         \gls{3N} forces without $\Delta$ excitations are included; in the 
         bottom panel, \gls{EDF}s derived from \gls{NN}+\gls{3N} forces with 
         $\Delta$ excitations are shown.
}
\label{fig:sp}
\end{figure}

The figure \ref{fig:sp} shows the example of neutron \gls{sp} states in 
$^{208}$Pb for the various \gls{EDF}s listed in Table \ref{tab:EDF}. Contrary 
to binding energies, we do not observe a very clear improvement or degradation 
of the shell structure as a function of the \gls{EDF} used. The 
\gls{sp} spectrum in other closed-shell nuclei yields similar conclusions. This 
could be attributed to the fact that the optimization protocol is the same for 
all \gls{EDF}s and explicitly include a constraint on a few spin-orbit 
splittings. In addition, work with either Skyrme \gls{EDF} or covariant 
\gls{DFT} suggest that correlations such as particle-vibration couplings play a 
major role in improving the shell structure in closed shell nuclei. It is  
unlikely that the \gls{DME} functionals we consider have this type of 
correlations built-in.

\subsection{Deformation Properties}

As mentioned in the introduction, one of the primary applications of \gls{DFT} 
is the study of nuclear structure and excited states for deformed nuclei, 
including rotational and vibrational spectra and fission. Although the shell 
structure of closed-shell nuclei discussed in the previous section and the 
large-scale mass tables shown in Figure \ref{fig:masses} are indicative of an 
\gls{EDF} with a good overall predictive power, it is important to also test 
the behavior of the \gls{EDF} at large deformations. For this reason, the 
{\UNEDFTWO} optimization protocol includes the excitation energy of a few 
fission isomers in actinides. As discussed in \cite{nikolov2011,
kortelainen2012}, this provides constraints on both the shell structure -- 
inasmuch as deformation properties are partly driven by the particular ordering 
and level density of \gls{sp} $j$-shells in spherical nuclei -- and on surface 
properties of the \gls{EDF}, which are related in particular to $\asym$.

\begin{table}[!htb]
\caption{Excitation energy of the fission isomer and height of the first and 
         second fission barriers in $^{240}$Pu for each of the 8 \gls{DME} 
         \gls{EDF}s, compared with empirical values in MeV. The column marked 
         $E_A$ (est.) includes an approximate 2.5 MeV correction on the height 
         of the first barrier to account for the fact that calculations were 
         done in axial symmetry.}
\begin{ruledtabular}
\begin{tabular}{ccccc}
		& $E_{\text{FI}}$ & $E_A$ & $E_A$ (est.) & $E_B$ \\
\hline
\gls{LO}                    &   2.625 & 9.479 & 6.979 & 6.468 \\
\gls{NLO}                   &   2.893 & 9.122 & 6.622 & 5.634 \\
\gls{N2LO}                  &   2.784 & 9.472 & 6.972 & 5.998 \\
\gls{N2LO}+\gls{3N}         &   2.481 & 8.992 & 6.492 & 6.955 \\
\gls{NLO}$\Delta$           &   2.395 &10.064 & 7.564 & 6.235 \\
\gls{NLO}$\Delta$+\gls{3N}  &   2.387 & 8.901 & 6.401 & 6.652 \\
\gls{N2LO}$\Delta$          &   2.691 & 9.967 & 7.467 & 7.214 \\
\gls{N2LO}$\Delta$+\gls{3N} &   2.350 &12.162 & 9.662 & 7.530 \\
Exp.                        &   2.800 &  -    & 6.050 & 5.150
\end{tabular}
\end{ruledtabular}
\label{table:barriers}
\end{table}

\begin{figure}[!htb]
\includegraphics[width=0.45\textwidth]{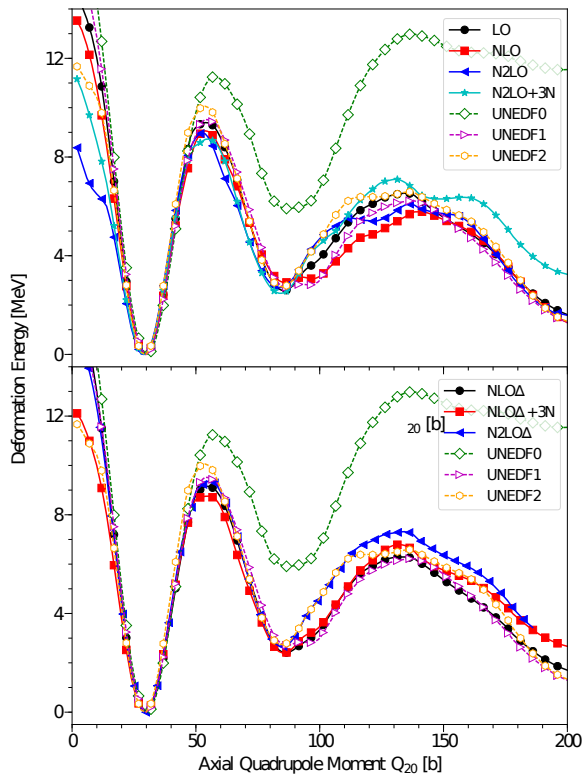}
\caption{Deformation potential energy surface in $^{240}$Pu as a function of 
         the axial quadrupole moment. Top panel: Energy functionals at 
         \gls{LO}, \gls{NLO}, \gls{N2LO}, with and without \gls{3N} forces 
         (when applicable). Bottom panel: Same for \gls{EDF} based on
         potentials including the $\Delta$ contribution. For comparison, each 
         panel also shows the results for the {\UNEDFZERO}, {\UNEDFONE}, and 
         {\UNEDFTWO} functionals of \cite{kortelainen2010,kortelainen2012,
         kortelainen2014}.}
\label{fig:fission}
\end{figure}

We report in Table \ref{table:barriers} the excitation energy of the fission 
isomer and height of the first and second barriers extracted from calculations 
of the potential energy curve in $^{240}$Pu. Across all 8 \gls{DME} \gls{EDF}s, 
the \gls{rms} deviation for the excitation energy of the fission isomer is 
0.29 MeV, which is comparable to the predictive power of the {\UNEDF} Skyrme 
functionals across all actinides; see \cite{kortelainen2014}. However, fission 
barriers tend to be too high -- even when taking into account the extra 
$\approx 2.5$ MeV caused by the lack of triaxiality in calculations of the 
first barrier, column marked $E_A$ (est.) in the table. For example, the 
\gls{rms} deviation for the second barrier is 1.55 MeV, compared with 1.39 MeV 
for {\UNEDFTWO} and 0.69 MeV for {\UNEDFONE} (across all actinides). We show in 
the two panels of Figure \ref{fig:fission} the potential energy curve of 
$^{240}$Pu as a function of the axial quadrupole moment for all the 8 
\gls{EDF}s considered here and listed in Table \ref{tab:EDF}, together with 
that of the {\UNEDFZERO}, {\UNEDFONE} and {\UNEDFTWO} functionals for 
comparison. 


\section{Conclusions}

In this work, we have calibrated and validated a set of energy density
functionals derived from local chiral potentials through the \gls{DME}.
We have provided the parametrization of these \gls{EDF}s at \gls{LO}, 
\gls{NLO}, \gls{N2LO}, \gls{N2LO}+\gls{3N}, \gls{NLO}$\Delta$, 
\gls{NLO}$\Delta$+\gls{3N}, \gls{N2LO}$\Delta$, \gls{N2LO}$\Delta$+\gls{3N} for 
a value of $R_c = 1.0$ fm and $n = 6$ in the regulator cutoff. The optimization 
was performed with the {\UNEDFTWO} protocol, and results were validated on the 
\gls{EOS} of infinite nuclear matter and pure neutron matter, nuclear mass 
tables, the shell structure of doubly-closed shell nuclei and the deformation 
energy of $^{240}$Pu.

The overall predictive power of these \gls{EDF}s is better than the Skyrme 
\gls{EDF}s obtained with the same optimization protocol. The relatively large 
variations among the 8 considered parametrizations is very encouraging as it 
suggests that even without ``beyond mean-field effects'' such as zero-point 
correlations energies, particle-vibration couplings, etc., observables such as 
binding energies are sensitive to the details of the \gls{EDF}. It is 
remarkable that, on average, the quality of the prediction increases noticeably 
as one goes further up in the chiral expansion. The exception is the 
\gls{N2LO}$\Delta$+\gls{3N} \gls{EDF}, for which predictions of binding 
energies and even the \gls{sp} structure degrade relatively to other 
\gls{EDF}s. However, one should keep in mind that, from a statistical 
perspective, calibrating functionals at the single-reference level implies that 
the model has ``defects'', i.e., it is not designed to accurately reproduce 
specific observables; see discussion on section \ref{sr-edf}. In practice, 
trying to fit both binding energies in closed shell nuclei like $^{208}$Pb and 
well-deformed nuclei in the rare-earth region could lead to overfitting issues. 

In this work, we have left out the estimate of uncertainties -- only 
quantifying numerical errors induced by approximating Yukawa form factors by a 
sum of Gaussian functions and by interpolating coupling functions. In 
particular, it could be worth studying in more details the exchange 
contribution to the energy by (i) calibrating a Hartree-only functional where 
the exchange contribution of the long-range chiral potential would be dropped 
entirely, and (ii) conversely, calibrating a functional where the Fock 
contribution from the chiral potential would be computed exactly by expanding 
it onto a sum of Gaussians, like the Hartree term. If we restrict ourselves to 
\gls{NN} potentials only, the computational effort is not significantly larger.

Since the \gls{DME} \gls{EDF}s originate from \gls{NN} and \gls{3N} potentials 
from chiral perturbation theory and their coupling constants have been 
determined with the exact same optimization protocol, these \gls{EDF} lend 
themselves particularly well to studies of systematic uncertainties. Together 
with the well-established machinery to quantify statistical uncertainties with 
either covariance or Bayesian methods, such studies could shed more light on 
the true predictive power of these \gls{EDF}s. Since the \gls{DME} functionals 
seem to encode some effects traditionally associated with beyond mean-field 
physics, it would also be natural to explore fits at the \gls{MR-EDF} level.


\begin{acknowledgments}
Support for this work was partly provided through the Scientific
Discovery through Advanced Computing (SciDAC) program funded by
U.S. Department of Energy, Office of Science, Advanced Scientific
Computing Research and Nuclear Physics. It was partly performed under
the auspices of the US Department of Energy by the Lawrence Livermore
National Laboratory under Contract DE-AC52-07NA27344. This work was
supported by the US Department of Energy under grant number
DE-FG02-93ER-40756 and the National Science Foundation under
grant number PHY--1614460. Computing support for this work came from the
Lawrence Livermore National Laboratory (LLNL) Institutional Computing
Grand Challenge program.

\end{acknowledgments} 

\bibliography{biblio,books,zotero_output}

\end{document}